\documentclass[aip,graphicx,amsmath,amssymb,app,,reprint,onecolumn]{revtex4-1}
\usepackage{graphicx}
\usepackage{amssymb}
\usepackage{bm}
\usepackage{textcomp}
\usepackage{xcolor}	
	
\begin{document}

	

\title{Elastic Wave Eigenmode Solver for Acoustic Waveguides}

\author{Nathan Dostart}
%\thanks{}
\affiliation{Department of Electrical, Computer, and Energy Engineering, University of Colorado, Boulder, CO 80309, USA}

\author{Yangyang Liu}
%\thanks{}
\affiliation{Department of Electrical, Computer, and Energy Engineering, University of Colorado, Boulder, CO 80309, USA}

\author{Milo\v s A. Popovi\'c}
\email[]{mpopovic@bu.edu}
%\thanks{}
\affiliation{Department of Electrical and Computer Engineering, Boston University, Boston, MA 02215, USA}

\date{\today}

\begin{abstract}
	 A numerical solver for the elastic wave eigenmodes in acoustic waveguides of inhomogeneous cross-section is presented. Operating under the assumptions of linear, isotropic materials, it utilizes a finite-difference method on a staggered grid to solve for the acoustic eigenmodes of the vector-field elastic wave equation. Free, fixed, symmetry, and anti-symmetry boundary conditions are implemented, enabling efficient simulation of acoustic structures with geometrical symmetries and terminations. Perfectly matched layers are also implemented, allowing for the simulation of radiative (leaky) modes. The method is analogous to eigenmode solvers ubiquitously employed in electromagnetics to find waveguide modes, and enables design of acoustic waveguides as well as seamless integration with electromagnetic solvers for optomechanical device design. The accuracy of the solver is demonstrated by calculating eigenfrequencies and mode shapes for common acoustic modes in several simple geometries and comparing the results to analytical solutions where available or to numerical solvers based on more computationally expensive methods.
\end{abstract}


\maketitle %\maketitle must follow title, authors, abstract and \pacs

\section{Introduction}
\label{sec:introduction}

Recent advances in several fields have attracted growing interest in the design of chip-scale acoustic devices that can interface with electrical and optical integrated components. Optomechanics is a prime example, where interacting acoustic and optical fields enable novel functionalities, such as ultra-sensitive quantum measurements \cite{anetsberger2010measuring,teufel2009nanomechanical}, narrow-linewidth lasers \cite{lee2012chemically,grudinin2009brillouin}, optomechanical memory \cite{bagheri2011dynamic,dong2015brillouin}, non-reciprocity and optical diodes \cite{poulton2012design,weis2010optomechanically,kim2015non}, optical cooling \cite{bahl2012observation,chan2011laser,teufel2011sideband}, phononic topological insulators \cite{peano2015topological,schmidt2015optomechanical}, optical amplifiers \cite{kittlaus2015large}, improved gravity wave detection \cite{arcizet2006radiation,kippenberg2008cavity}, microwave filters \cite{byrnes2012photonic}, and quantum state transfer \cite{verhagen2012quantum,palomaki2013coherent,bochmann2013nanomechanical}. The field of RF micro-electromechanical systems is another important example where electroacoustic transduction of bulk and surface acoustic waves in acoustic resonators enables some devices which outperform conventional RF electronics. These include reconfigurable filters \cite{lu2013reconfigurable}, narrowband signal filtering \cite{psychogiou2015acoustic}, and high quality factor (Q) resonators \cite{clark2005high}. These devices are also being integrated into microelectronic systems for improved performance \cite{weinstein2010resonant}. For all these acoustic wave based devices, good performance requires confining the acoustic energy to a small cross-sectional area (waveguides) or volume (resonators), phase-matching the acoustic wave to transducer arrays and/or optical waves, and optimizing transduction efficiency. Numerical tools for designing and simulating acoustic waveguide modes are thus necessary to enable efficient device designs, intricate nanoscale coupling schemes, and novel device architectures.

Previous work has predominantly focused on full-wave simulation of 2D and 3D domains due to the importance of these problems in geophysics. 3D solvers have been developed for anisotropic, heterogeneous domains using both finite-difference \cite{petersson2015wave} and finite-element \cite{gao2015generalized} methods and are currently the predominant method for designing acoustic devices in the GHz frequency range. While many commercial software tools \cite{multiphysics2012comsol} allow the design of acoustic waveguides using a full 3D solver, sometimes more efficiently by reducing the volume using Floquet (periodic) boundary conditions, the most efficient approach is the maximally reduced, 2D problem formulation (disregarding cross-section symmetries). The 2D formulation returns an orthogonal, complete set of modes (field and frequency) at a specified propagation constant. In contrast, the solution of a 3D volume returns all resonant modes including those in higher order Brillouin zones, which are an artifact of the 3D formulation and usually undesired. This rigorous reduction of a 3D problem to a 2D simulation domain is referred to as a 2+1D simulation. Papers utilizing a 2+1D version of the finite-element method have been implemented to solve free, isotropic waveguide geometries in the ultrasound regime \cite{gavric1995computation} and extended to axially symmetric waveguides \cite{wilcox2002dispersion}, embedded waveguides \cite{castaings2008finite}, and viscoelastic materials \cite{bartoli2006modeling}. It should be noted that generic FEM tools such as COMSOL, which provide an interface to solve arbitrary partial differential equations, are in principle capable of solving the 2+1D Cartesian acoustic waveguide simulation (by entering custom equations).

In this work, we demonstrate an acoustic waveguide mode solver based on the finite-difference method, analogous to electromagnetic (EM) mode solvers, which solves the linear isotropic elastic wave equation. The mode solver combines all capabilities of FEM mode solvers with the efficiency of finite-differencing, all physical boundary conditions as well as perfectly matched layers, directly overlaps with the optical Yee grid~\cite{yee1966numerical}, and is verified for acoustic frequencies ranging from $1\,$MHz to $10\,$GHz. We utilize a staggered-grid discretization that preserves second-order accuracy for all physical quantities of interest, by analogy with the Yee scheme first used in EM~\cite{yee1966numerical} and identical to the staggered grid of \cite{madariaga1976dynamics,etemadsaeed2016no}. The solver finds the acoustic eigenmodes of a structure that is invariant along one Cartesian dimension. The source-free vectorial elastic wave equation in an inhomogeneous, isotropic, linear medium is formulated as an eigenvalue equation. Given a specific material configuration and the propagation constant, a unique set of eigenmodes with corresponding modal frequencies can be found. The acoustic problem is only a linear eigenvalue problem when formulated such that the propagation constant $\beta$ is given and the eigenfrequency $\omega$ is the eigenvalue. This is notably different than the analogous electromagnetic problem which, due to Gauss's Law, can be formulated as a linear eigenvalue problem with either $\omega$ or $\beta$ given and the other variable as the eigenvalue. A finite-difference method is used to sample the inhomogeneous medium and acoustic field over the computational domain, the cross-section of the structure. This operation transforms the continuous eigenvalue problem into a sparse matrix which can be solved by standard numerical sparse matrix eigen solver methods (e.g. shifted inverse method via Arnoldi iteration and a sparse linear solver). An example case is the `eigs' function in MATLAB~\cite{matlab}. The solver calculates, to second order accuracy on the finite-difference grid, the acoustic eigenmodes of a straight acoustic waveguide with specified cross-section and propagation constant. The use of 2+1D mode solvers in both optical and acoustic domains allows for accurate and efficient calculation of propagation parameters and coupling terms, which can be input to the acousto-optic simulations for accuracy comparable to full 3D simulations. To this end, we have made our mode solver code, in MATLAB implementation, freely available \cite{mathworks}.

We present the theoretical and mathematical basis for the elastic wave equation eigenmode decomposition in Sec.~\ref{sec:theory}. This section also includes implementation details such as the finite-differencing scheme and boundary conditions. We then implement the method in MATLAB and validate its accuracy by analysis of the matrix construction, convergence tests, example cross-sections with analytical solutions, and other examples which can be solved numerically in 3D using a commercial FEM solver \cite{multiphysics2012comsol} in Sec.~\ref{sec:validation}.

\section{Theory, Mathematical Framework, and Implementation of the Mode Solver}
\label{sec:theory}

The mathematical basis for the eigenmode solver and its implementation are described in this section. The source-free, linear, isotropic elastic wave equation is cast as an eigenmode problem with frequency eigenvalues and discretized. The only assumptions made are that the cross-section is invariant along one linear direction, $\hat{z}$; all materials are linear, isotropic, and time-invariant; and that a computational domain that is finite in cross-section represents well the modes of the structure (justified for `confined' modes, such as in electromagnetics~\cite{snyder2012optical}). The isotropic assumption is made for convenience and is not essential; the method can be applied to anisotropic media. This approach to solving for waveguide modes is analogous to electromagnetic mode solvers, where instead of an electric or magnetic field we solve for the elastic displacement field. Notably, in the elastic equation, there is no equivalent of Gauss's Law and thus we solve for three field components and have three polarization families. In EM, Gauss's Law reduces the eigen problem to two polarization mode families and specification of two field components fully defines the mode (e.g. $E_x$, $E_y$).

\subsection{Derivation of Isotropic Linear Elastic Wave Equation}
\label{subsec:derivation}

We begin with Newton's 2\textsuperscript{nd} law written in a density formulation, the strain-displacement relation, and generalized Hooke's law
\vspace{-4pt}
\begin{equation}
\label{eq:Newt}
\rho\partial_t^2\mathbf{u}=\nabla\cdot\bar{\bar{\sigma}}
\end{equation}
\vspace{-10pt}
\begin{equation}
\label{eq:strain_disp}
\bar{\bar{\epsilon}}=\nabla_s\mathbf{u}
\end{equation}
\vspace{-10pt}
\begin{equation}
\label{eq:hooke}
\bar{\bar{\sigma}}=\mathbb{C}:\bar{\bar{\epsilon}}
\end{equation}
where $\rho(\mathbf{r},t)\equiv\rho(\mathbf{r})$ is the spatial distribution of material density, $\mathbf{u}(\mathbf{r},t)$ is the elastic displacement field, $\nabla\cdot$ is the tensor divergence, $\bar{\bar{\sigma}}(\mathbf{r},t)$ is the stress tensor, $\bar{\bar{\epsilon}}(\mathbf{r},t)$ is the strain tensor, $\nabla_s$ is the symmetric spatial vector gradient, $\mathbb{C}(\mathbf{r},t)\equiv\mathbb{C}(\mathbf{r})$ is the fourth order stiffness tensor, and $:$ denotes a tensor inner product. Note that while stress and strain are second order tensors, due to symmetry considerations they can be unwrapped as six-vectors following the Voigt notation ($\bar{\bar{\sigma}}=\left[\sigma_{xx}\,\sigma_{yy}\,\sigma_{zz}\,\sigma_{yz}\,\sigma_{xz}\,\sigma_{xy}\right]^T$) \cite{Auld1990}, which is the form used in this paper. The strain tensor has an analogous form. The operator $\nabla_s$, which acts on a vector to give a rank 2 tensor, is the adjoint of the tensor divergence operator ($-\nabla\cdot$). $\mathbb{C}$ is a symmetric operator (even in the presence of loss) due to our assumption of linear, time-invariant materials \cite{Auld1990}, and we use the notation $\partial_i$ to refer to the partial derivative $\partial/\partial i$.

Substituting Eqs.~\eqref{eq:strain_disp}-\eqref{eq:hooke} into Eq.~\eqref{eq:Newt} in order to factor out the stress tensor yields the linear elastic wave equation written in terms of the displacement field
\vspace{-4pt}
\begin{equation}
\label{eq:WaveEq}
\rho\partial_t^2\mathbf{u}=\nabla\cdot \mathbb{C}:\nabla_s\mathbf{u}.
\end{equation}

We define a weighted displacement field $\tilde{\mathbf{u}}\equiv\sqrt{\rho}\mathbf{u}$. This allows Eq.~\eqref{eq:WaveEq}, a generalized eigenvalue problem, to be recast as an ordinary eigenvalue problem with a Hermitian operator in the absence of loss. Invariance of $\rho$ and $\mathbb{C}$ with time means that the system has a spectrum defined by a linear eigenvalue problem by setting $\partial_t \rightarrow j\omega$. The elastic wave equation can be written as
\vspace{-4pt}
\begin{equation}
\label{eq:EigEq}
\omega^2\tilde{\mathbf{u}}=\frac{-1}{\sqrt{\rho}}\nabla\cdot \mathbb{C}:\nabla_s\frac{1}{\sqrt{\rho}}\tilde{\mathbf{u}}.
\end{equation}

This has the form of an eigenvalue equation ($\Lambda x=\bar{\bar{H}} x$), so we identify the eigenvalue as $\Lambda \equiv \omega^2$, the eigenvector as $x\equiv\tilde{\mathbf{u}}$, and the symmetrized elastic resonance operator $\bar{\bar{H}}$ as
\vspace{-4pt}
\begin{equation}
\label{eq:EigMat}
\bar{\bar{H}} = \frac{-1}{\sqrt{\rho}}\nabla\cdot \mathbb{C}:\nabla_s\frac{1}{\sqrt{\rho}}.
\end{equation}

To clarify this expression, the modified differential operators (in the form that operate on the reduced, six-vector notation used for $\bar{\bar{\sigma}}$ and $\bar{\bar{\epsilon}}$) can be written in matrix form
\vspace{-4pt}
\begin{equation}
\label{eq:nablas}
\nabla_s=\begin{bmatrix}
\partial_x & 0 & 0 \\
0 & \partial_y & 0 \\
0 & 0 & \partial_z \\
0 & \partial_z & \partial_y \\
\partial_z & 0 & \partial_x \\
\partial_y & \partial_x & 0 \\
\end{bmatrix}
\end{equation}
\vspace{-4pt}
\begin{equation}
\label{eq:div}
\nabla\cdot=\begin{bmatrix}
\partial_x & 0 & 0 & 0 & \partial_z & \partial_y \\
0 & \partial_y & 0 & \partial_z & 0 & \partial_x \\
0 & 0 & \partial_z & \partial_y & \partial_x & 0 \\
\end{bmatrix}.
\end{equation}

To this point, we have defined the acoustic resonator problem (in 3D). Next, when solving for eigenmodes of a structure with $z$-invariant geometry, by Fourier transformation along $z$ the $z$-directed derivative becomes the propagation constant ($\partial_z=-j\beta$). This reduces the problem to one on the cross-sectional plane and ensures that only modes with the specified propagation constant will be returned by the solver.

Next, making the isotropic assumption, the stiffness tensor (in the form that operates on the six-vector notation) can be reduced to
\vspace{-4pt}
\begin{equation}
\label{eq:C}
\mathbb{C}=\begin{bmatrix}
\lambda+2\mu & \lambda & \lambda & 0 & 0 & 0 \\
\lambda & \lambda+2\mu & \lambda & 0 & 0 & 0 \\
\lambda & \lambda & \lambda+2\mu & 0 & 0 & 0 \\
0 & 0 & 0 & \mu & 0 & 0 \\
0 & 0 & 0 & 0 & \mu & 0 \\
0 & 0 & 0 & 0 & 0 & \mu \\
\end{bmatrix}
\end{equation}
where $\lambda(\mathbf{r},t)\equiv\lambda(\mathbf{r})$ and $\mu(\mathbf{r},t)\equiv\mu(\mathbf{r})$ are the first and second Lam\'{e} parameters.

\subsection{Discretization Scheme}
\label{subsec:discretization}

\begin{figure}[t]
	\centering
	\includegraphics[width=.6\textwidth]{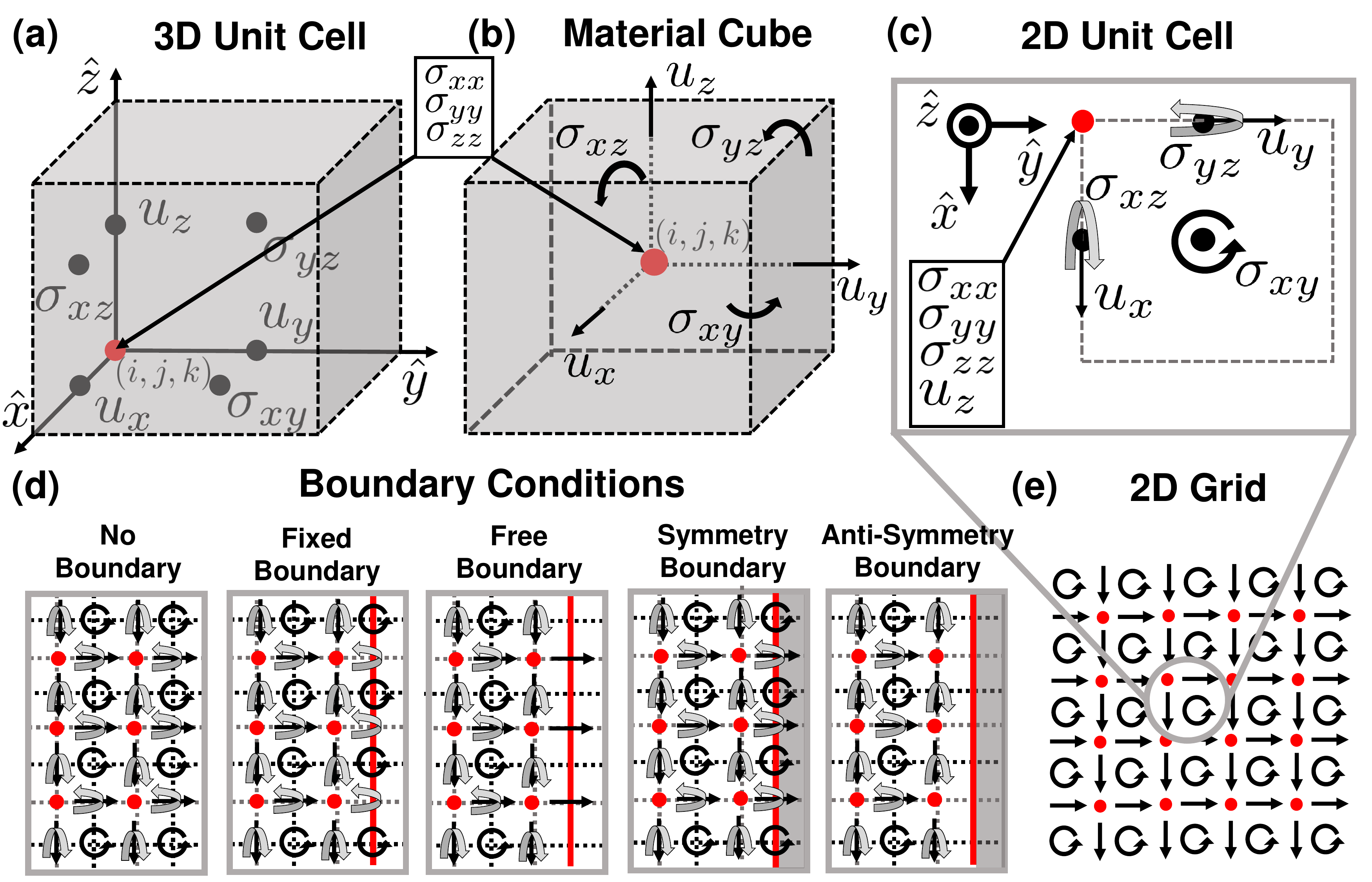}
	\caption{(a) Unit cell of the 3D discretization grid used in this paper, where the unit cell coordinates $(i,j,k)$ define the corner of the unit cell and are co-located with the principal stresses. Each point is denoted with the quantities which are sampled at that point.(b) Discrete `material cube' which provides the basis for an intuitive choice of discretization grid. The `material cube' is offset from the unit cell by a half-step along each dimension. (c) 2D unit cell which can be obtained by collapsing the 3D unit cell along the $z$-dimension. (d) Representative implementation of the boundary conditions, including choice of location (red line) and values sampled on the boundary (or removed). (e) 2D discretization grid with schematic depiction of component grid locations.}
	\vspace{-15pt}
	\label{fig:grid}
\end{figure}

In order to solve the eigenvalue problem in an arbitrary cross-section geometry ($\rho$, $\mathbb{C}$) numerically, Eq.~\eqref{eq:EigEq} is discretized to arrive at a form with a finite number of degrees of freedom, and is cast as a matrix eigenvalue problem. An appropriate 3D grid is first formulated which accurately captures the physics to second-order accuracy in the discretization and the grid is then collapsed to the 2D simulation domain. The collapse is performed such that the cross-sectional locations of all grid points are unchanged while the $z$-coordinates of all grid points are set to a single value. The $z$-derivatives, which would be performed between two points in the 3D case, amount to multiplying a single grid point by $-j\beta$ in the 2D case. More formally, for a 3D grid with $z$-invariance, $\partial_z\rightarrow -j\beta \mathrm{sin}(\beta\Delta z/2)/(\beta\Delta z/2) \exp(-j\beta\Delta z/2)$~\cite{chew1994em}, but as $\Delta z\rightarrow0$ then $\partial_z\rightarrow -j\beta$.

The discretization grid used is depicted in Fig.~\ref{fig:grid}. This grid is based on physical intuition from solid mechanics: the state of an element cube is primarily described by the principal stresses ($\sigma_{xx}$, $\sigma_{yy}$, $\sigma_{zz}$), which we define to reside at the center of the cube. The principal stresses lead to the deformation of the faces of the cube, such that the center of each face is the sampling location of the associated normal displacement. If grid point $(i,j,k)$ is associated with the principal stresses (and the center of the cubic volume elements), the corresponding displacements are $u_x:(i+1/2,j,k)$, $u_y:(i,j+1/2,k)$, and $u_z:(i,j,k+1/2)$. An appropriate grid for the shear stresses is found to be a further half-step offset, such that shear stresses are located at $\sigma_{xy}: (i+1/2,j+1/2,k)$, $\sigma_{xz}: (i+1/2,j,k+1/2)$, and $\sigma_{yz}:(i,j+1/2,k+1/2)$. The strain grid is co-located with the stress grid. The finite-difference operators that approximate the spatial partial derivatives transform quantities from the stress/strain grid to the displacement grid and back, as expected and desired. Because isotropic materials are assumed, the material operator $\mathbb{C}$ does not induce a change of grid coordinates (the strain at a specific grid point is only related to the stresses at the same grid point). This method can also be extended to anisotropic materials at the cost of additional complexity in the $\mathbb{C}$ matrix, which must then have different coefficients sampled on different grids.

The staggered-grid scheme used here is analogous to the Yee grid \cite{yee1966numerical}, commonly used in EM solvers, which preserves second-order accuracy for all fields due to centered differencing (when all materials/coefficients vary spatially on the scale of the discretization). Our discretization grid is slightly more complicated in that the elastic displacement field and shear stress directly replicate a Yee grid, while the principal stresses occupy an additional position at the center of each `cube'.

\subsection{Finite Difference Operators}
\label{subsec:fd_operators}

Centered differences in 2\textsuperscript{nd} order differential equations can be formed by appropriate combinations of forward and backward differences on a staggered grid \cite{chew1994em}. Denoting forward differences as $\tilde{\partial_i}$ and backward differences as $\hat{\partial_i}$, the differential operators can be rewritten in terms of forward and backward differences (where the propagation constant has been substituted for $\partial_z$) as
\vspace{-4pt}
\begin{equation}
\label{eq:nablas_fd}
\nabla_s=\begin{bmatrix}
\hat{\partial_x} & 0 & 0 \\
0 & \hat{\partial_y} & 0 \\
0 & 0 & -j\beta \\
0 & -j\beta & \tilde{\partial_y} \\
-j\beta & 0 & \tilde{\partial_x} \\
\tilde{\partial_y} & \tilde{\partial_x} & 0 \\
\end{bmatrix}
\end{equation}
\vspace{-4pt}
\begin{equation}
\label{eq:div_fd}
\nabla\cdot=\begin{bmatrix}
\tilde{\partial_x} & 0 & 0 & 0 & -j\beta & \hat{\partial_y} \\
0 & \tilde{\partial_y} & 0 & -j\beta & 0 & \hat{\partial_x} \\
0 & 0 & -j\beta & \hat{\partial_y} & \hat{\partial_x} & 0 \\
\end{bmatrix}.
\end{equation}
It should be noted that, because $\tilde{\partial_i}^\dagger=-\hat{\partial_i}$, $\nabla_s$ and $-\nabla\cdot$ remain adjoints in discrete form ($\nabla_s^\dagger=-\nabla\cdot$).

\subsection{Matrix Operator Construction}
\label{subsec:tm_constr}

Referring to Eq.~\eqref{eq:EigMat}, the resonance operator can be rewritten as a matrix. This matrix can be found by discretizing the displacement field and unwrapping the field into a single column vector. A convenient ordering was found as
\vspace{-4pt}
\begin{equation}
\arraycolsep=0pt
\label{eq:unwrap_u}
\tilde{\mathbf{u}}=\begin{bmatrix}
\left\{u_x\right\}, & \left\{u_y\right\}, & \left\{u_z\right\}
\end{bmatrix}^T
\end{equation}
\vspace{-15pt}
\begin{equation}
\label{eq:unwrap_ux}
\{u_x\}=\begin{bmatrix}
u_x^{(0,0)} & u_x^{(1,0)} & \cdots & u_x^{(n^{u_x}_x,0)} & u_x^{(0,1)} & u_x^{(1,1)} & \cdots & u_x^{(n^{u_x}_x,1)} & \cdots & u_x^{(n^{u_x}_x,n^{u_x}_y)} 
\end{bmatrix}^T.
\end{equation}

Thus, each vector component is unwrapped in the $x-y$ plane along the $x$-dimension first and the components are then concatenated. This results in a vector of approximate length $3n_xn_y$ from an $n_x\times n_y\times 3$ matrix. A similar method is used to unwrap the stress tensor in a vector of approximate length $6n_xn_y$ of the form
\vspace{-4pt}
\begin{equation}
\label{eq:unwrap_stress}
\bar{\bar{\sigma}}=\begin{bmatrix}
\left\{\sigma_{xx}\right\} & \left\{\sigma_{yy}\right\} & \left\{\sigma_{zz}\right\} & \left\{\sigma_{yz}\right\} & \left\{\sigma_{xz}\right\} & \left\{\sigma_{xy}\right\}
\end{bmatrix}^T
\end{equation}
with the strain tensor having an equivalent form. Note that the different components $\sigma_{ij}$ and $u_i$ are of different lengths since they are sampled at different locations. Shown in Table~\ref{tb:grid} are the number of grid points in both the $x$- and $y$-directions ($n^{comp}_x$, $n^{comp}_y$) where the length of the vectorized form of each component is simply $n^{comp}_{vec}=n_x^{comp} n_y^{comp}$.

\begin{table}[h!]
	\centering
	\begin{tabular}{ |c|c|c|c|c|c|c|c|c|c| }
		\hline
		Component & $u_x$ & $u_y$ & $u_z$ & $\sigma_{xx}$ & $\sigma_{yy}$ & $\sigma_{zz}$ & $\sigma_{yz}$ & $\sigma_{xz}$ & $\sigma_{xy}$\\
		\hline
		$n^{comp}_x$ & $n_x+1$ & $n_x$ & $n_x$ & $n_x$ & $n_x$ & $n_x$ & $n_x$ & $n_x+1$ & $n_x+1$\\
		\hline
		$n^{comp}_y$ & $n_y$ & $n_y+1$ & $n_y$ & $n_y$ & $n_y$ & $n_y$ & $n_y+1$ & $n_y$ & $n_y+1$\\
		\hline
	\end{tabular}
	\caption{Grid points of each elastic wave component.}
	\vspace{-10pt}
	\label{tb:grid}
\end{table}

The differential operator hereby becomes a \textit{matrix} operator where each entry in Eqs.~\eqref{eq:C}, \eqref{eq:nablas_fd}, \eqref{eq:div_fd} is now a block matrix which relates one component to another. After populating the $\nabla\cdot$, $\mathbb{C}$, and $\nabla_s$ matrices with the block matrices, they are multiplied together to give
\vspace{-4pt}
\begin{equation}
\label{eq:matrix_simp}
\bar{\bar{H}}=\frac{-1}{\sqrt{\rho}}\begin{bmatrix}
\tilde{\partial}_xc_{11}\hat{\partial}_x-\mu\beta^2+\hat{\partial}_y\mu\tilde{\partial}_y & \tilde{\partial}_x\lambda\hat{\partial}_y+\hat{\partial}_y\mu\tilde{\partial}_x & -j\beta\tilde{\partial}_x\lambda-j\beta\mu\tilde{\partial}_x \\
\tilde{\partial}_y\lambda\hat{\partial}_x+\hat{\partial}_x\mu\tilde{\partial}_y & \tilde{\partial}_yc_{11}\hat{\partial}_y-\mu\beta^2+\hat{\partial}_x\mu\tilde{\partial}_x & -j\beta\tilde{\partial}_y\lambda-j\beta\mu\tilde{\partial}_y \\
-j\beta\lambda\hat{\partial}_x-j\beta\hat{\partial}_x\mu & -j\beta\lambda\hat{\partial}_y-j\beta\hat{\partial}_y\mu & -\beta^2c_{11}+\hat{\partial}_y\mu\tilde{\partial}_y+\hat{\partial}_x\mu\tilde{\partial}_x
\end{bmatrix}\frac{1}{\sqrt{\rho}}
\end{equation}
where we have shortened $\lambda+2\mu$ to the Voigt notation $c_{11}$ \cite{Auld1990}.

We here note that the matrix, currently in a Hermitian form in the absence of loss, can be cast as a real symmetric matrix $\bar{\bar{H}}^\prime$ (a special case of Hermitian matrices) by defining a modified eigenvector form $\tilde{\mathbf{u}}^\prime\equiv\left[\tilde{u}_x,\,\tilde{u}_y,\,-j\tilde{u}_z\right]$, i.e. we expect that the $u_z$ component will be in quadrature with the $u_x$, $u_y$ components. Hermitian matrices have real eigenvalues (energy conservation) and their eigenmodes form a complete, orthogonal set. Casting the operator matrix into a symmetric form, which has eigenvectors that can be chosen to be entirely real, additionally implies that $u_z$ is guaranteed to be in quadrature with $u_x$ and $u_y$. If loss is present, then this modified matrix form will be complex symmetric rather than real symmetric and we would expect to obtain complex eigenvalues and non-orthogonal eigenmodes. Defining the modified operator matrix $\bar{\bar{H}}^\prime$ such that $\omega^2\tilde{\mathbf{u}}^\prime=\bar{\bar{H}}^\prime\tilde{\mathbf{u}}^\prime$, we can calculate this modified operator matrix by recognizing that $\tilde{\mathbf{u}}^\prime=\bar{\bar{R}}\tilde{\mathbf{u}}$ and $\bar{\bar{H}}^\prime=\bar{\bar{R}}\bar{\bar{H}}\bar{\bar{R}}^{-1}$ where $\bar{\bar{R}}$ is a diagonal matrix with $\mathrm{diag}(\bar{\bar{R}}) = [1,\,1,\,-j]$. We can then write the modified operator matrix as
\vspace{-4pt}
\begin{equation}
\label{eq:matrix_mod}
\bar{\bar{H}}^\prime=\frac{-1}{\sqrt{\rho}}\begin{bmatrix}
\tilde{\partial}_xc_{11}\hat{\partial}_x-\mu\beta^2+\hat{\partial}_y\mu\tilde{\partial}_y & \tilde{\partial}_x\lambda\hat{\partial}_y+\hat{\partial}_y\mu\tilde{\partial}_x & -\beta\tilde{\partial}_x\lambda-\beta\mu\tilde{\partial}_x \\
\tilde{\partial}_y\lambda\hat{\partial}_x+\hat{\partial}_x\mu\tilde{\partial}_y & \tilde{\partial}_yc_{11}\hat{\partial}_y-\mu\beta^2+\hat{\partial}_x\mu\tilde{\partial}_x & -\beta\tilde{\partial}_y\lambda-\beta\mu\tilde{\partial}_y \\
\beta\lambda\hat{\partial}_x+\beta\hat{\partial}_x\mu & \beta\lambda\hat{\partial}_y+\beta\hat{\partial}_y\mu & -\beta^2c_{11}+\hat{\partial}_y\mu\tilde{\partial}_y+\hat{\partial}_x\mu\tilde{\partial}_x
\end{bmatrix}\frac{1}{\sqrt{\rho}}.
\end{equation}

We remind the reader here that, even for complex finite difference operators, $\hat{\partial}_i^T=-\tilde{\partial}_i$ so that this modified matrix $\bar{\bar{H}}^\prime$ is symmetric. It is additionally real if both material parameters ($\lambda$, $\mu$) and finite difference operators ($\hat{\partial}_i$, $\tilde{\partial}_i$) are real, corresponding to no material loss and no complex coordinate stretching radiation (absorbing boundaries), respectively.

\subsection{Modal Orthogonality}
\label{subsec:orth}

The orthogonality condition for vectors (in the sense of 1D tensors) is $\mathbf{x}_i^\dagger \mathbf{x}_j\equiv |\mathbf{x}_i|^2\delta_{ij}$ where the vector norm is defined as $|\mathbf{x}_i|\equiv (\mathbf{x}_i^\dagger \mathbf{x}_i)^{1/2}$. These two equations correspond to $\int\int \mathbf{x}_i(\mathbf{r}_T)^*\cdot\mathbf{x}_j(\mathbf{r}_T)dA=|\mathbf{x}_i||\mathbf{x}_j|\delta_{ij}$ and $|\mathbf{x}_i|=(\iint \mathbf{x}_i(\mathbf{r}_T)^*\cdot\mathbf{x}_i(\mathbf{r}_T)dA)^{1/2}$ when $\mathbf{x}$ represents a 3-vector field in a 2D domain. Noting that that $\tilde{\mathbf{u}}^{\prime*}_i\cdot\tilde{\mathbf{u}}^\prime_j=\rho\mathbf{u}^*_i\cdot\mathbf{u}_j$, we can then replace $\mathbf{x}\rightarrow\tilde{\mathbf{u}}$ and write the orthogonality condition and normalization as
\vspace{-4pt}
\begin{equation}
\label{eq:orth}
\iint\rho\mathbf{u}_i^*\cdot\mathbf{u}_jdA=|\tilde{\mathbf{u}}_i||\tilde{\mathbf{u}}_j|\delta_{ij}
\end{equation}
\vspace{-4pt}
\begin{equation}
\label{eq:norm}
|\tilde{\mathbf{u}}_i|=\left(\iint\rho\mathbf{u}_i^*\cdot\mathbf{u}_idA\right)^{\frac{1}{2}}.
\end{equation}

Noting that the particle velocity field is $\mathbf{v}_i\equiv\partial_t\mathbf{u}_i=j\omega_i\mathbf{u}_i$, we can substitute $\mathbf{u}_i=-j\mathbf{v}_i/\omega_i$ into Eqs.~\eqref{eq:orth}-\eqref{eq:norm} to obtain measures of the kinetic energy $\left(E_K=\rho|\mathbf{v}|^2/2\right)$. Alternatively, because the potential energy $\left(V=\bar{\bar{\epsilon}}^\dagger\mathbb{C}\bar{\bar{\epsilon}}/2\right)$ is equivalent to the kinetic energy when averaged \cite{Auld1990}, this gives the orthogonality condition that would have been obtained if the wave equation were formulated with $\bar{\bar{\epsilon}}$ as the free variable. This is analogous to the electromagnetic case, where resonator problems use an energy formulation to determine modal orthogonality \cite{joannopoulos2011photonic}.

\subsection{Boundary Conditions}
\label{subsec:bound_cond}

Four boundary conditions are implemented: fixed boundary ($\mathbf{u}|_{Bound}=0$), free boundary ($\bar{\bar{\sigma}}\cdot\hat{n}=0$), symmetry boundary ($\partial_nu_n=0$, $\mathbf{u}_t|_{Bound}=0$), and anti-symmetry boundary ($u_n|_{Bound}=0$, $\partial_n\mathbf{u}_t=0$). Perfectly matched layers (PMLs), radiation-absorbing regions which permit simulation of radiating structures \cite{chew1996perfectly}, are also implemented. For all boundary conditions, we locate the boundary such that the boundary-normal displacements ($u_n$), corresponding out-of-plane shear stresses ($\sigma_{nz}$), and in-plane shear stresses ($\sigma_{xy}$) lie on the boundary. The choice of boundary location is depicted in Fig.~\ref{fig:grid}(d) as the location of the red line. We additionally allow for vacuum (no material) in the simulation domain.

The fixed boundary condition requires the displacement field be set to zero on the boundary. An efficient and computationally stable implementation is to remove the corresponding boundary-normal displacements $u_n$ (since they are the only ones coincident with the boundary) from the matrix operator and solution vector, which is the method used here. We validate this boundary condition in Sec.~\ref{subsec:valid_num} by confirming that a beam with fixed boundary conditions returns the correct eigenmode shapes and frequencies (Fig.~\ref{fig:freefixbeam}).

For the free surface boundary condition, the boundary-normal stress components must go to zero at the boundary. In a similar manner to the implementation of the fixed boundary condition, we remove the shear stresses $\sigma_{nz}$, $\sigma_{xy}$ on the boundary. This boundary condition is also validated in Sec.~\ref{subsec:valid_num} by confirming that the eigenmode shapes and frequencies are accurately calculated (Fig.~\ref{fig:freefixbeam}).

Symmetry and anti-symmetry boundary conditions are hybrids of free and fixed boundary conditions. Similar to the case in electromagnetics, symmetry of the normal displacement corresponds to anti-symmetry of the transverse displacement, while anti-symmetry of the normal displacement corresponds to symmetry of the transverse displacement. A symmetry (anti-symmetry) boundary therefore refers to symmetry (anti-symmetry) of the normal displacement and anti-symmetry (symmetry) of the transverse displacement.

The symmetry and anti-symmetry boundary conditions are implemented by combinations of the free and fixed boundary conditions. The proper implementation of these boundary conditions is validated in Fig.~\ref{fig:symmbc} by comparison of the free beam modes calculated with and without the symmetry/anti-symmetry boundary conditions.

We enable vacuum within the simulation domain by removing both stress and displacement grid points within the vacuum region. We also remove the stress sampled on the vacuum-material boundary and double boundary-crossing derivatives as appropriate. This exactly emulates a free surface boundary condition along the interface, as desired.

We additionally implement PMLs in order to support simulation of leaky modes. The PML is not formally a boundary condition and is instead a region within the computational domain \cite{chew1996perfectly,chew19943d,berenger1994perfectly,teixeira1998analytical}. A desired use for this acoustic mode solver is to estimate propagation losses of a particular acoustic mode in a waveguide, and a key loss mechanism can be radiation loss (analogous to a distributed version of clamping loss in resonators, e.g. cantilevers). We refer to acoustic modes exhibiting radiation loss as a leaky modes, which can still be considered confined modes when the radiation loss rate is much smaller than the oscillation frequency.

The PML boundary condition can be implemented as a complex coordinate stretching near the boundary \cite{teixeira1998analytical}. This can be achieved by making the grid discretization components $\Delta x$ and $\Delta y$ complex within the PML region. We choose to use a fixed boundary condition adjacent to the PML in order to force the acoustic field to go to zero at the boundary. In order to replicate the PML used by the commercial solver used for comparison, a linear coordinate stretching was used of the form
\vspace{-4pt}
\begin{equation}
\label{eq:PML}
\Delta x_{PML} = \Delta x\left(1-j\alpha\left|\frac{x-x_{PML,begin}}{x_{PML,end}-x_{PML,begin}}\right|\right)\quad x_{PML,end}\geq x \geq x_{PML,begin}
\end{equation}
where $\alpha$ is the scaling parameter of the PML. To demonstrate the accuracy of the PMLs implemented in this solver, we constructed a leaky acoustic waveguide that tunnels radiation across a barrier into the continuum on both sides (Fig.~\ref{fig:leakywg}).

\section{Mode Solver Validation and Demonstration}
\label{sec:validation}

In order to demonstrate the accuracy of the proposed acoustic mode solver, we examine the resonance matrix to verify its Hermitian nature in Sec.~\ref{subsec:valid_math}, then compare several simple cases in our MATLAB implementation to both analytical solutions (Sec.~\ref{subsec:valid_theory}) and a commercial, numerical 3D solver (Sec.~\ref{subsec:valid_num}). We verify the symmetry and anti-symmetry conditions in Sec.~\ref{subsec:symmbc} and demonstrate support of leaky modes in Sec.~\ref{subsec:demo}.

\subsection{Mathematical Validation: Matrix Properties}
\label{subsec:valid_math}

\begin{figure}[t]
	\centering
	\includegraphics[width=.6\textwidth]{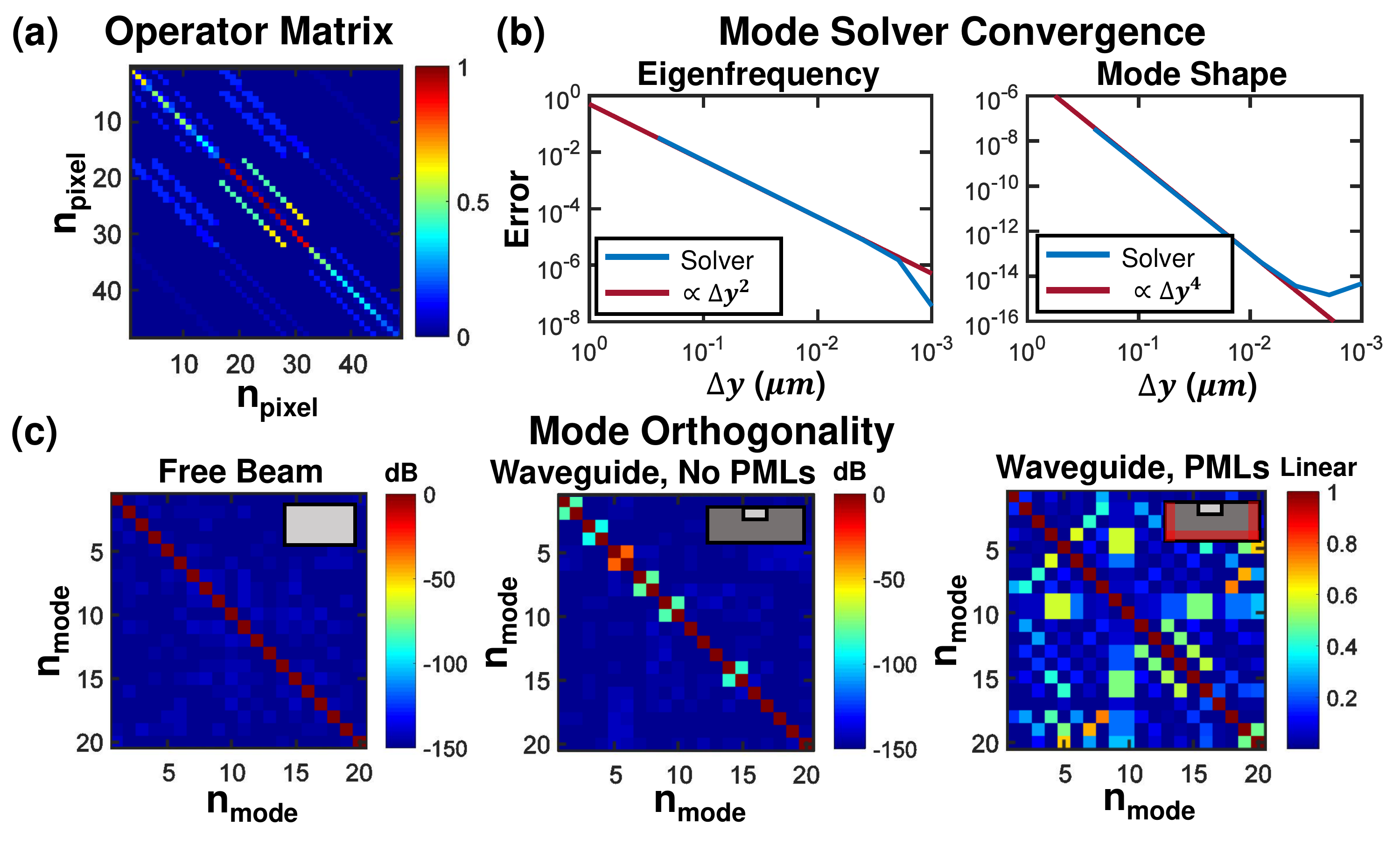}
	\caption{Validation of resonance operator matrix properties. (a) Example of the operator matrix. (b) Convergence tests of the mode solver: the $20\,$dB/decade slope for eigenfrequency error ($40\,$dB/decade for mode shape) indicates second-order accuracy. (c) Tests of modal orthogonality in three configurations, supporting the Hermitian nature of the resonance operator in the absence of PMLs (and non-Hermitian nature with PMLs).}
	\vspace{-15pt}
	\label{fig:math_val}
\end{figure}

To test our discretization scheme, we examine the properties of resonance operator matrices generated by the mode solver code. To provide an example of the resonance operator matrix, we implement a trial cross-section with two free and two fixed boundary conditions on a $4\times4$ pixel grid in Fig.~\ref{fig:math_val}(a). We additionally verify in Fig.~\ref{fig:math_val}(b) second order scaling of the error with discretization. For this case, we chose an anti-symmetric Lamb wave with a $10\,${\textmu}m wavelength in a $1\,${\textmu}m thick free slab and varied the discretization along the slab thickness direction. The calculated eigenfrequency and mode shape errors are relative to the analytical solutions, described further in Sec.~\ref{subsec:valid_theory}. Our convergence plots indicate that the mode solver eigenfrequency error scales as $\varepsilon\propto (\Delta y)^2$, i.e. that our mode solver is second-order accurate as expected. The mode shape error scales as $\varepsilon\propto (\Delta y)^4$ due to the squaring in the mode shape error formula. We additionally expect that, in the absence of material loss or PMLs, the resonance operator would be real and symmetric even after discretization which we verified using a $200\times200$ pixel grid simulation with both free and fixed boundary conditions. We also confirmed that the operator is complex symmetric when PMLs are used. Other hallmarks of Hermitian operators are real eigenvalues, orthogonal eigenvectors, and that the eigenvectors form a complete basis. We have shown an example demonstrating the orthogonality of the eigenvectors in both a free beam and a waveguide without PMLs, and lack thereof when PMLs are implemented, in Fig.~\ref{fig:math_val}(c).

\begin{figure}[t]
	\centering
	\includegraphics[width=.7\textwidth]{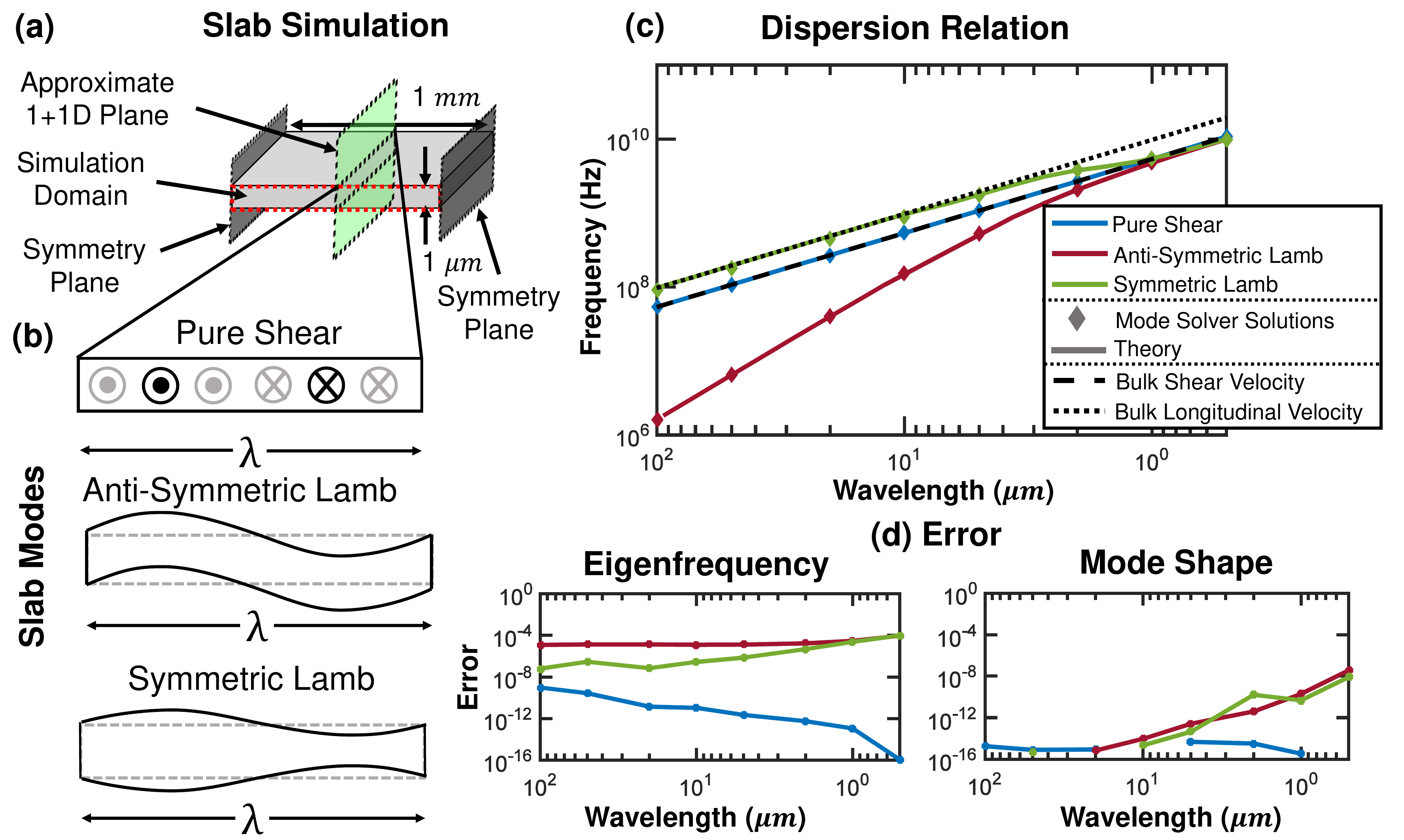}
	\caption{Three acoustic modes (pure shear, anti-symmetric Lamb, and symmetric Lamb) in a $1\,${\textmu}m thick silicon slab were simulated and compared to the theoretical modal distributions and frequencies: (a) Configuration schematic. (b) Illustration of the three acoustic modes simulated. (c) Modal dispersion curves; solid lines are theoretical dispersions and diamond markers are mode solver solutions (shear and longitudinal bulk wave asymptotes also shown). (d) Relative error in frequency and mode shape $\left(\varepsilon_{MS}=1-(\left|\int U_{MS}\cdot U_{o}^*\mathrm{dx}\right|^2)/(\int\left| U_{MS} \right|^2\mathrm{dx}\int\left| U_{o} \right|^2\mathrm{dx})\right)$ between the mode solver and theory. The pure shear wave mode shape error is near the level of numerical error ($\sim10^{-14}$) and is therefore not shown.}
	\label{fig:slab}
	\vspace{-15pt}
\end{figure}

\subsection{Analytical Validation: Free Slab Waveguide}
\label{subsec:valid_theory}

We choose a slab waveguide, which can be analytically solved, to compare the proposed mode solver with elastic wave theory. The simulation implementation and results are shown in Fig.~\ref{fig:slab}. We investigate three fundamental acoustic waves in a thin plate [Fig.~\ref{fig:slab}(b)], namely pure shear waves, anti-symmetric Lamb waves, and symmetric Lamb waves \cite{Auld1990}. We calculate the theoretical dispersion relations [Fig.~\ref{fig:slab}(c)] of these waves compared with the mode solver solutions as the wavelength is varied from 100$\,${\textmu}m to 0.5$\,${\textmu}m to achieve full coverage of the thin plate regime and into the thick plate (infinite half-space) regime. For the mode solver solutions, we use a plate which is much wider than its thickness ($1\times1000\,${\textmu}m) with the appropriate symmetry boundary conditions on the edges to approximate an infinitely long (1D cross-section) plate. We then calculate the error of the mode solver relative to theory, both in terms of mode shape and eigenfrequency, in Fig.~\ref{fig:slab}(d). We sample the acoustic mode distribution at the center of the plate as an approximation to the theoretical 1+1D case. The mode shape error used here is formulated as $\varepsilon_{MS}=1-(\left|\iint U_{MS}\cdot U_{o}^*\mathrm{dA}\right|^2)/(\iint\left| U_{MS} \right|^2\mathrm{dA}\iint\left| U_{o} \right|^2\mathrm{dA})$ where $U_o$ is the true mode shape and $U_{MS}$ is the mode shape found by the solver. For the slab case, we integrate over only a single dimension as the mode shapes are invariant along the second cross-sectional dimension. For the general case, the integration is carried out over the entire cross-section. The mode solver reproduces the key aspects of these acoustic waves with $<1\%$ error across a range of wavelengths which captures the thick, thin, and wavelength-scale plate thickness regimes, indicating that the mode solver faithfully models linear elastic physics.

\begin{figure}[t]
	\centering
	\includegraphics[width=\textwidth]{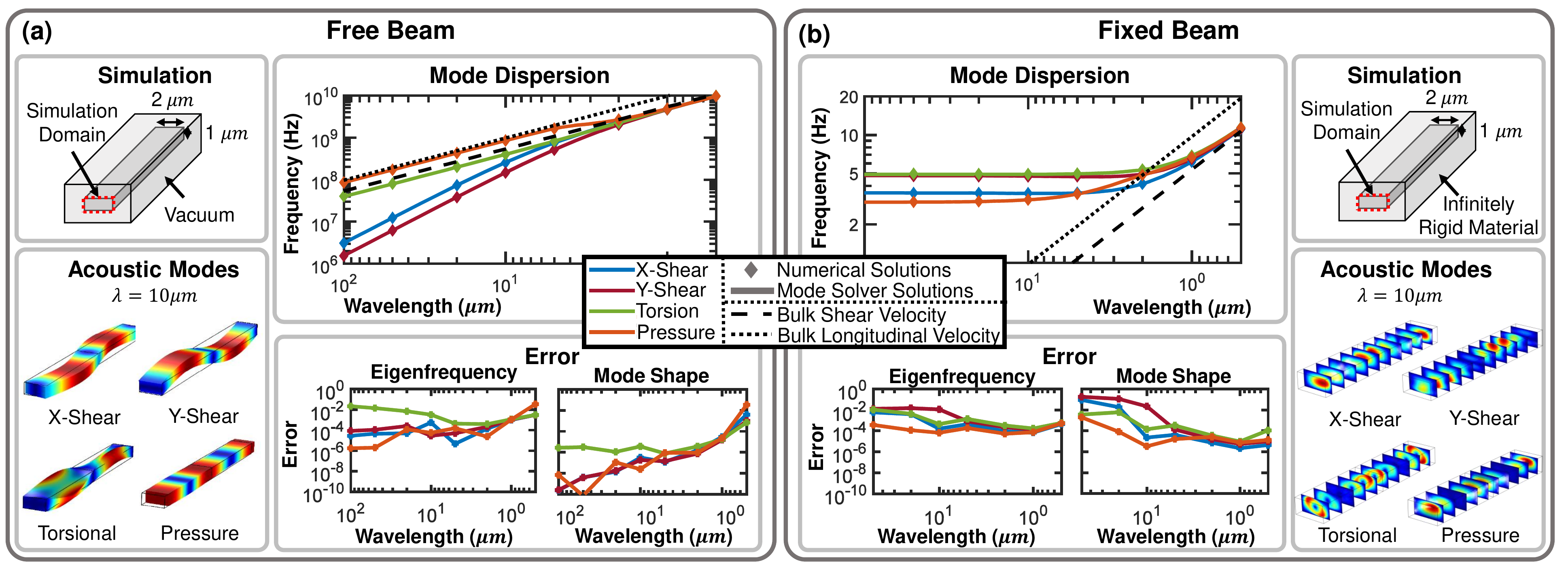}
	\caption{Four acoustic modes ($x$- and $y$-shear, torsion, and pressure waves) in suspended (a) and fixed (b) silicon beams were simulated across a wavelength range of $100\,${\textmu}m to $0.5\,${\textmu}m in both the mode solver and the 3D solver. Shown in each box for the suspended and fixed configurations are a schematic of the simulation implementation, depictions from the 3D solver of the acoustic modes, modal dispersions, and relative error between the mode solver and the numerical solver. The solid lines in the dispersion plots are the mode solver solutions and the diamond markers are the numerical solver solutions.}
	\vspace{-15pt}
	\label{fig:freefixbeam}
\end{figure}

\subsection{Numerical Validation: Free and Fixed Beam Waveguides}
\label{subsec:valid_num}

We now consider a rectangular beam waveguide (in both suspended and fixed, i.e. encased in an infinitely rigid medium, configurations) to compare the proposed mode solver with numerical solutions obtained from a commercial 3D solver \cite{multiphysics2012comsol}. We investigate the lowest order mode in each of four basic mode families: vertically ($y$) polarized shear waves, horizontally ($x$) polarized shear waves, torsional waves, and pressure waves. We choose as our cross-section a $2\times 1\,${\textmu}m silicon beam so as to avoid a degeneracy of the $x$- and $y$-shear modes. The same geometry is implemented in the commercial 3D solver with the beam length chosen equal to the simulated wavelength and Floquet periodic boundary conditions applied to the $z$-oriented boundaries. Both solvers are then used to find these four acoustic modes and corresponding eigenfrequencies across a wavelength range of $100\,${\textmu}m to $0.5\,${\textmu}m. These comparisons are made in Fig.~\ref{fig:freefixbeam} where we depict the comparative error of the mode solver relative to the 3D solver in terms of frequency and modeshape. For the majority of wavelengths, both the mode shape and frequency are very close ($<1\%$) to the 3D solver results which are defined here as the reference value.

\subsection{Symmetry/Anti-symmetry Boundary Condition Validation}
\label{subsec:symmbc}

\begin{figure}[t]
	\centering
	\includegraphics[width=.7\textwidth]{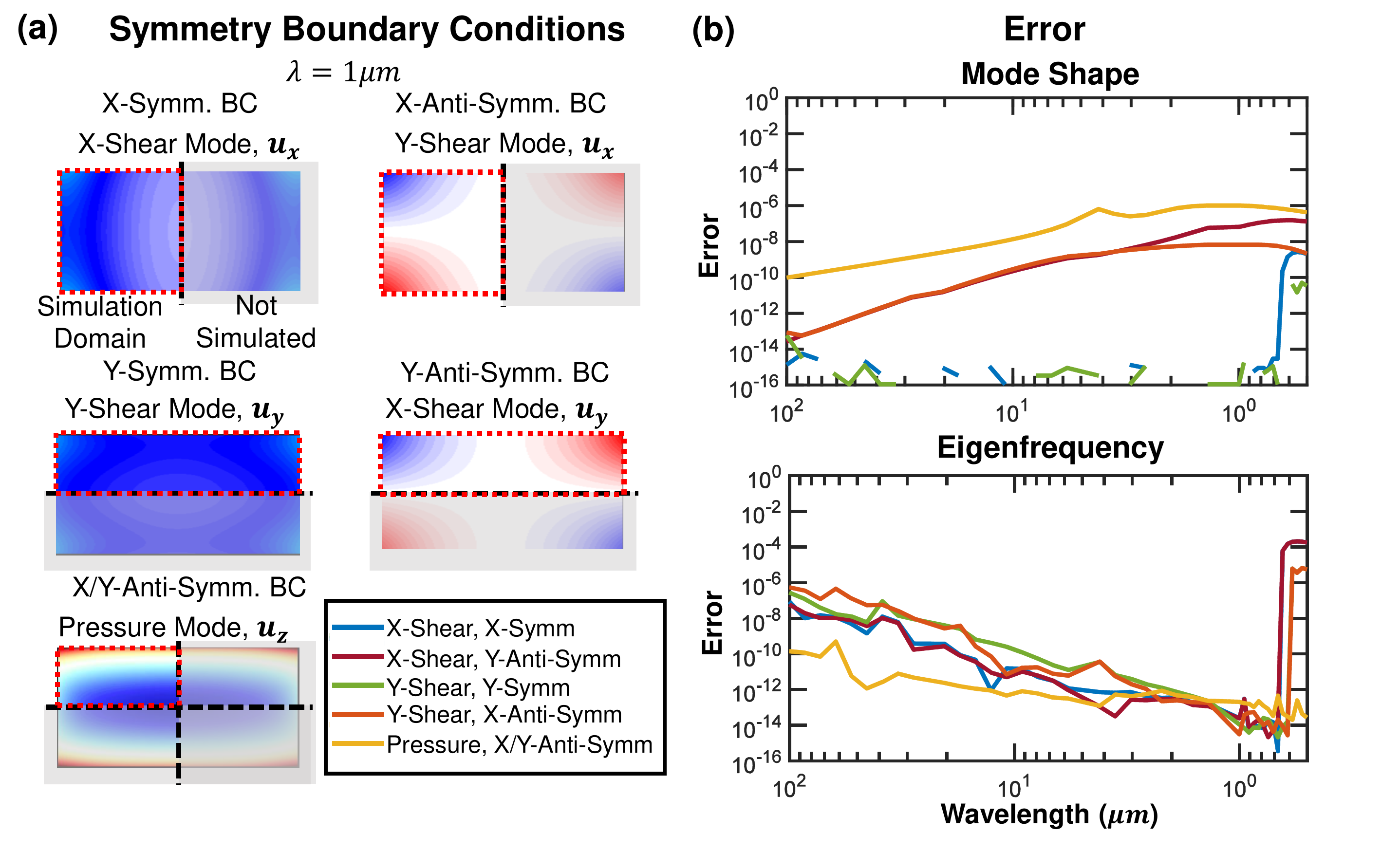}
	\caption{Three acoustic modes ($x$- and $y$-shear and pressure waves) are simulated in a suspended silicon beam in a truncated simulation domain using symmetric and anti-symmetric boundary conditions and compared with the same modes simulated in the full simulation domain. (a) Schematic depictions of the symmetric and anti-symmetric boundary conditions and simulated mode shapes. (b) Comparison of mode shape and eigenfrequency error between the modes obtained with the full simulation domain, and the truncated simulation domain using symmetry/anti-symmetry boundary conditions. Increased errors at the shortest wavelengths are due to the mode solver returning mixed modes when the eigenfrequencies are nearly degenerate.}
	\vspace{-15pt}
	\label{fig:symmbc}
\end{figure}

Having demonstrated the accuracy of the solver, we can use the free beam modes as a basis for comparison with modes computed in a reduced simulation domain that takes advantage of geometrical symmetries. For this case we halve (or quarter) the simulation domain and use a symmetry/anti-symmetry boundary condition on the cut plane to recreate the same mode. We have chosen to use both boundary conditions to recreate the same mode as a redundant test of the boundary conditions, and the pressure mode is chosen as a demonstration that two orthogonal boundary conditions can be used at the same time without loss of accuracy. This comparison is shown in Fig.~\ref{fig:symmbc}, where examples of the modes and symmetry conditions are shown in Fig.~\ref{fig:symmbc}(a) and the errors are shown in Fig.~\ref{fig:symmbc}(b). The jump in error in both eigenfrequency and mode shape at small wavelengths is due to the simulated modes being nearly degenerate in frequency at these wavelengths, causing the solver to return mixed mode shapes and eigenfrequencies when the full simulation domain is used. The halved/quartered simulation domains did not have mixing due to the imposed symmetry requirements.

\subsection{Solver Demonstration: Evanescent Waveguiding, Leaky Modes, and PMLs}
\label{subsec:demo}

\begin{figure}[t]
	\centering
	\includegraphics[width=.95\textwidth]{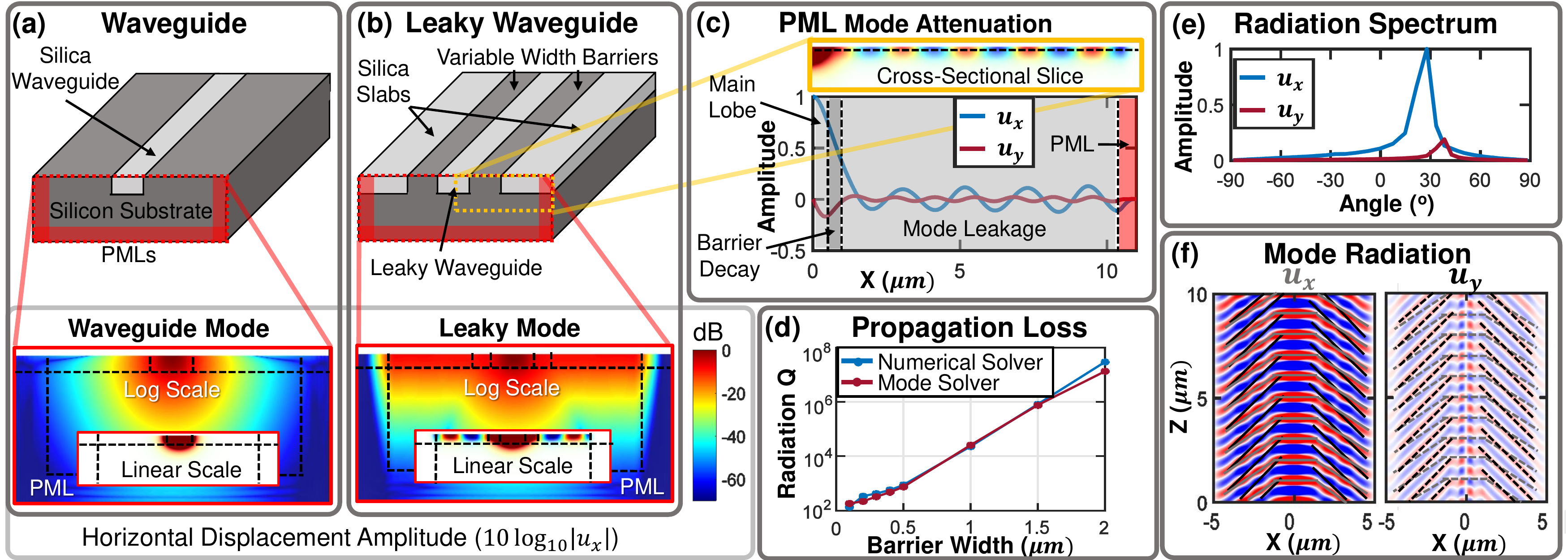}
	\caption{An embedded strip acoustic waveguide ($1\,\times0.4\,${\textmu}m) is simulated to demonstrate evanescent confinement, wave-guiding, radiation tunneling, leaky modes, and PML operation. (a) Schematic depiction of an acoustic waveguide and log-scale, cross-sectional plot of the horizontal displacement of the quasi-Love wave guided mode with linear scale inset. (b) The same waveguide after silica slabs have been added (with a $0.5\,${\textmu}m silicon barrier) with plots of the leaky mode. (c) An extended 1D slice of the leaky acoustic mode in (b) showing the displacement amplitude. (d) Propagation loss comparison between the 3D solver and the mode solver. (e) Fourier transform of displacement amplitudes in the leaky section of the 1D slice from (c), demonstrating coupling into two radiation modes. (f) Mode propagation and radiated wavefronts (shown schematically) for both horizontal and vertical displacement components.}
	\vspace{-15pt}
	\label{fig:leakywg}
\end{figure}

We now construct an acoustic waveguide formed of two materials (a `core' and a `cladding') and find evanescently confined acoustic modes. Adding a radiation layer (formed of core material) to the waveguide cross-section creates a leaky acoustic waveguide that tunnels radiation across a (cladding) barrier into the radiation-mode continuum on both sides (see analogous optical slab waveguide \cite{popovic2008theory}). Implementation of PMLs on the simulation domain edges then allows outward radiating waves to be absorbed without reflection, enabling the simulation of leaky modes analogous to electromagnetic solvers~\cite{feng2002computation} and calculation of waveguide mode radiation losses.

The configuration, modal amplitude plots, radiation patterns, and calculated radiation losses are shown in Fig.~\ref{fig:leakywg}. First, an evanescently guiding waveguide is constructed [Fig.~\ref{fig:leakywg}(a)] by embedding a strip of silica within a silicon substrate. This waveguide supports two `polarizations' of acoustic modes, analogous to the Love and Rayleigh surface waves \cite{Auld1990}. The former wave is chosen and the depicted log-scale plot of the horizontal displacement amplitude demonstrates that the wave is evanescently confined. We then add silica slabs to either side of the silica waveguide, separated by variable width silicon barriers, which cause the waveguide and guided acoustic modes to couple to the slab radiation continuum on both sides and become `leaky' [Fig.~\ref{fig:leakywg}(b)]. Oscillation of the field within the silica layer can be seen in the inset, and non-decaying intensity within the leaky region can be seen in the log-scale plot; both are indicative of the radiative loss of the guided mode into the adjacent silica layers.

An extended simulation domain with a longer silica layer further demonstrates the presence of this mode leakage in Fig.~\ref{fig:leakywg}(c) where a cross-sectional slice plots both horizontal and vertical displacement amplitudes within the different material regions. The mode displacement is predominantly confined in the waveguide with an evanescent tail in the silicon barrier. Some of the acoustic energy tunnels across the barrier into the silica slab, where it becomes oscillatory, and then enters the PML and is attenuated as expected. A comparative plot of this propagation loss, as a function of silicon barrier width, is shown in Fig.~\ref{fig:leakywg}(d) as numerical validation of the accuracy of the PML. By taking a Fourier transform of the mode leakage section we show that two primary radiation field components are excited in the leaky wave [Fig.~\ref{fig:leakywg}(e)], corresponding to quasi-Love and quasi-Rayleigh waves in the silica slab. Fig.~\ref{fig:leakywg}(f) shows a cross-section in the $x-z$ plane depicting the horizontal (left) and vertical (right) displacement amplitudes, where the wavefronts have been outlined for both polarizations and overlaid on both plots (gray for horizontal displacement, black for vertical).

\section{Conclusion}
\label{sec:conclusion}
We have developed an eigenmode solver based on a finite-difference scheme on a staggered-grid for the linear, isotropic elastic wave equation. We have verified the accuracy of the solver by comparison with both theory and a 3D numerical solver (COMSOL). We have also used these comparisons to demonstrate correct implementation of fixed and free boundary conditions. We then used a leaky acoustic waveguide to demonstrate the solver's ability to simulate evanescent guiding, leaky modes, and numerically accurate PMLs. We expect that the mode solver will be of utility in the design of on-chip acoustic wave devices. It is particularly suited to interfacing with electromagnetic solvers on the Yee grid for applications based on acousto-optics and optomechanics.

\begin{acknowledgments}
The authors acknowledge helpful discussions with Yossef Erhlichman, Cale Gentry, and Bohan Zhang. This work was supported by a National Science Foundation Graduate Research Fellowship (Grant \#1144083) and a 2012 Packard Fellowship for Science and Engineering (Grant \#2012-38222).
\end{acknowledgments}

%\bibliography{biblio}

\begin{thebibliography}{47}%
\makeatletter
\providecommand \@ifxundefined [1]{%
 \@ifx{#1\undefined}
}%
\providecommand \@ifnum [1]{%
 \ifnum #1\expandafter \@firstoftwo
 \else \expandafter \@secondoftwo
 \fi
}%
\providecommand \@ifx [1]{%
 \ifx #1\expandafter \@firstoftwo
 \else \expandafter \@secondoftwo
 \fi
}%
\providecommand \natexlab [1]{#1}%
\providecommand \enquote  [1]{``#1''}%
\providecommand \bibnamefont  [1]{#1}%
\providecommand \bibfnamefont [1]{#1}%
\providecommand \citenamefont [1]{#1}%
\providecommand \href@noop [0]{\@secondoftwo}%
\providecommand \href [0]{\begingroup \@sanitize@url \@href}%
\providecommand \@href[1]{\@@startlink{#1}\@@href}%
\providecommand \@@href[1]{\endgroup#1\@@endlink}%
\providecommand \@sanitize@url [0]{\catcode `\\12\catcode `\$12\catcode
  `\&12\catcode `\#12\catcode `\^12\catcode `\_12\catcode `\%12\relax}%
\providecommand \@@startlink[1]{}%
\providecommand \@@endlink[0]{}%
\providecommand \url  [0]{\begingroup\@sanitize@url \@url }%
\providecommand \@url [1]{\endgroup\@href {#1}{\urlprefix }}%
\providecommand \urlprefix  [0]{URL }%
\providecommand \Eprint [0]{\href }%
\providecommand \doibase [0]{http://dx.doi.org/}%
\providecommand \selectlanguage [0]{\@gobble}%
\providecommand \bibinfo  [0]{\@secondoftwo}%
\providecommand \bibfield  [0]{\@secondoftwo}%
\providecommand \translation [1]{[#1]}%
\providecommand \BibitemOpen [0]{}%
\providecommand \bibitemStop [0]{}%
\providecommand \bibitemNoStop [0]{.\EOS\space}%
\providecommand \EOS [0]{\spacefactor3000\relax}%
\providecommand \BibitemShut  [1]{\csname bibitem#1\endcsname}%
\let\auto@bib@innerbib\@empty
%</preamble>
\bibitem [{\citenamefont {Anetsberger}\ \emph {et~al.}(2010)\citenamefont
  {Anetsberger}, \citenamefont {Gavartin}, \citenamefont {Arcizet},
  \citenamefont {Unterreithmeier}, \citenamefont {Weig}, \citenamefont
  {Gorodetsky}, \citenamefont {Kotthaus},\ and\ \citenamefont
  {Kippenberg}}]{anetsberger2010measuring}%
  \BibitemOpen
  \bibfield  {author} {\bibinfo {author} {\bibfnamefont {G.}~\bibnamefont
  {Anetsberger}}, \bibinfo {author} {\bibfnamefont {E.}~\bibnamefont
  {Gavartin}}, \bibinfo {author} {\bibfnamefont {O.}~\bibnamefont {Arcizet}},
  \bibinfo {author} {\bibfnamefont {Q.~P.}\ \bibnamefont {Unterreithmeier}},
  \bibinfo {author} {\bibfnamefont {E.~M.}\ \bibnamefont {Weig}}, \bibinfo
  {author} {\bibfnamefont {M.~L.}\ \bibnamefont {Gorodetsky}}, \bibinfo
  {author} {\bibfnamefont {J.~P.}\ \bibnamefont {Kotthaus}}, \ and\ \bibinfo
  {author} {\bibfnamefont {T.~J.}\ \bibnamefont {Kippenberg}},\ }\bibfield
  {title} {\enquote {\bibinfo {title} {Measuring nanomechanical motion with an
  imprecision below the standard quantum limit},}\ }\href@noop {} {\bibfield
  {journal} {\bibinfo  {journal} {Physical Review A}\ }\textbf {\bibinfo
  {volume} {82}},\ \bibinfo {pages} {061804} (\bibinfo {year}
  {2010})}\BibitemShut {NoStop}%
\bibitem [{\citenamefont {Teufel}\ \emph {et~al.}(2009)\citenamefont {Teufel},
  \citenamefont {Donner}, \citenamefont {Castellanos-Beltran}, \citenamefont
  {Harlow},\ and\ \citenamefont {Lehnert}}]{teufel2009nanomechanical}%
  \BibitemOpen
  \bibfield  {author} {\bibinfo {author} {\bibfnamefont {J.}~\bibnamefont
  {Teufel}}, \bibinfo {author} {\bibfnamefont {T.}~\bibnamefont {Donner}},
  \bibinfo {author} {\bibfnamefont {M.}~\bibnamefont {Castellanos-Beltran}},
  \bibinfo {author} {\bibfnamefont {J.}~\bibnamefont {Harlow}}, \ and\ \bibinfo
  {author} {\bibfnamefont {K.}~\bibnamefont {Lehnert}},\ }\bibfield  {title}
  {\enquote {\bibinfo {title} {Nanomechanical motion measured with an
  imprecision below that at the standard quantum limit},}\ }\href@noop {}
  {\bibfield  {journal} {\bibinfo  {journal} {Nature Nanotechnology}\ }\textbf
  {\bibinfo {volume} {4}},\ \bibinfo {pages} {820--823} (\bibinfo {year}
  {2009})}\BibitemShut {NoStop}%
\bibitem [{\citenamefont {Lee}\ \emph {et~al.}(2012)\citenamefont {Lee},
  \citenamefont {Chen}, \citenamefont {Li}, \citenamefont {Yang}, \citenamefont
  {Jeon}, \citenamefont {Painter},\ and\ \citenamefont
  {Vahala}}]{lee2012chemically}%
  \BibitemOpen
  \bibfield  {author} {\bibinfo {author} {\bibfnamefont {H.}~\bibnamefont
  {Lee}}, \bibinfo {author} {\bibfnamefont {T.}~\bibnamefont {Chen}}, \bibinfo
  {author} {\bibfnamefont {J.}~\bibnamefont {Li}}, \bibinfo {author}
  {\bibfnamefont {K.~Y.}\ \bibnamefont {Yang}}, \bibinfo {author}
  {\bibfnamefont {S.}~\bibnamefont {Jeon}}, \bibinfo {author} {\bibfnamefont
  {O.}~\bibnamefont {Painter}}, \ and\ \bibinfo {author} {\bibfnamefont
  {K.~J.}\ \bibnamefont {Vahala}},\ }\bibfield  {title} {\enquote {\bibinfo
  {title} {Chemically etched ultrahigh-{Q} wedge-resonator on a silicon
  chip},}\ }\href@noop {} {\bibfield  {journal} {\bibinfo  {journal} {Nature
  Photonics}\ }\textbf {\bibinfo {volume} {6}},\ \bibinfo {pages} {369--373}
  (\bibinfo {year} {2012})}\BibitemShut {NoStop}%
\bibitem [{\citenamefont {Grudinin}, \citenamefont {Matsko},\ and\
  \citenamefont {Maleki}(2009)}]{grudinin2009brillouin}%
  \BibitemOpen
  \bibfield  {author} {\bibinfo {author} {\bibfnamefont {I.~S.}\ \bibnamefont
  {Grudinin}}, \bibinfo {author} {\bibfnamefont {A.~B.}\ \bibnamefont
  {Matsko}}, \ and\ \bibinfo {author} {\bibfnamefont {L.}~\bibnamefont
  {Maleki}},\ }\bibfield  {title} {\enquote {\bibinfo {title} {Brillouin lasing
  with a {CaF2} whispering gallery mode resonator},}\ }\href {\doibase
  10.1103/PhysRevLett.102.043902} {\bibfield  {journal} {\bibinfo  {journal}
  {Phys. Rev. Lett.}\ }\textbf {\bibinfo {volume} {102}},\ \bibinfo {pages}
  {043902} (\bibinfo {year} {2009})}\BibitemShut {NoStop}%
\bibitem [{\citenamefont {Bagheri}\ \emph {et~al.}(2011)\citenamefont
  {Bagheri}, \citenamefont {Poot}, \citenamefont {Li}, \citenamefont
  {Pernice},\ and\ \citenamefont {Tang}}]{bagheri2011dynamic}%
  \BibitemOpen
  \bibfield  {author} {\bibinfo {author} {\bibfnamefont {M.}~\bibnamefont
  {Bagheri}}, \bibinfo {author} {\bibfnamefont {M.}~\bibnamefont {Poot}},
  \bibinfo {author} {\bibfnamefont {M.}~\bibnamefont {Li}}, \bibinfo {author}
  {\bibfnamefont {W.~P.}\ \bibnamefont {Pernice}}, \ and\ \bibinfo {author}
  {\bibfnamefont {H.~X.}\ \bibnamefont {Tang}},\ }\bibfield  {title} {\enquote
  {\bibinfo {title} {Dynamic manipulation of nanomechanical resonators in the
  high-amplitude regime and non-volatile mechanical memory operation},}\
  }\href@noop {} {\bibfield  {journal} {\bibinfo  {journal} {Nature
  Nanotechnology}\ }\textbf {\bibinfo {volume} {6}},\ \bibinfo {pages}
  {726--732} (\bibinfo {year} {2011})}\BibitemShut {NoStop}%
\bibitem [{\citenamefont {Dong}\ \emph {et~al.}(2015)\citenamefont {Dong},
  \citenamefont {Shen}, \citenamefont {Zou}, \citenamefont {Zhang},
  \citenamefont {Fu},\ and\ \citenamefont {Guo}}]{dong2015brillouin}%
  \BibitemOpen
  \bibfield  {author} {\bibinfo {author} {\bibfnamefont {C.-H.}\ \bibnamefont
  {Dong}}, \bibinfo {author} {\bibfnamefont {Z.}~\bibnamefont {Shen}}, \bibinfo
  {author} {\bibfnamefont {C.-L.}\ \bibnamefont {Zou}}, \bibinfo {author}
  {\bibfnamefont {Y.-L.}\ \bibnamefont {Zhang}}, \bibinfo {author}
  {\bibfnamefont {W.}~\bibnamefont {Fu}}, \ and\ \bibinfo {author}
  {\bibfnamefont {G.-C.}\ \bibnamefont {Guo}},\ }\bibfield  {title} {\enquote
  {\bibinfo {title} {Brillouin-scattering-induced transparency and
  non-reciprocal light storage},}\ }\href@noop {} {\bibfield  {journal}
  {\bibinfo  {journal} {Nature Communications}\ }\textbf {\bibinfo {volume}
  {6}} (\bibinfo {year} {2015})}\BibitemShut {NoStop}%
\bibitem [{\citenamefont {Poulton}\ \emph {et~al.}(2012)\citenamefont
  {Poulton}, \citenamefont {Pant}, \citenamefont {Byrnes}, \citenamefont {Fan},
  \citenamefont {Steel},\ and\ \citenamefont {Eggleton}}]{poulton2012design}%
  \BibitemOpen
  \bibfield  {author} {\bibinfo {author} {\bibfnamefont {C.~G.}\ \bibnamefont
  {Poulton}}, \bibinfo {author} {\bibfnamefont {R.}~\bibnamefont {Pant}},
  \bibinfo {author} {\bibfnamefont {A.}~\bibnamefont {Byrnes}}, \bibinfo
  {author} {\bibfnamefont {S.}~\bibnamefont {Fan}}, \bibinfo {author}
  {\bibfnamefont {M.}~\bibnamefont {Steel}}, \ and\ \bibinfo {author}
  {\bibfnamefont {B.~J.}\ \bibnamefont {Eggleton}},\ }\bibfield  {title}
  {\enquote {\bibinfo {title} {Design for broadband on-chip isolator using
  stimulated {B}rillouin scattering in dispersion-engineered chalcogenide
  waveguides},}\ }\href@noop {} {\bibfield  {journal} {\bibinfo  {journal}
  {Optics Express}\ }\textbf {\bibinfo {volume} {20}},\ \bibinfo {pages}
  {21235--21246} (\bibinfo {year} {2012})}\BibitemShut {NoStop}%
\bibitem [{\citenamefont {Weis}\ \emph {et~al.}(2010)\citenamefont {Weis},
  \citenamefont {Rivi{\`e}re}, \citenamefont {Del{\'e}glise}, \citenamefont
  {Gavartin}, \citenamefont {Arcizet}, \citenamefont {Schliesser},\ and\
  \citenamefont {Kippenberg}}]{weis2010optomechanically}%
  \BibitemOpen
  \bibfield  {author} {\bibinfo {author} {\bibfnamefont {S.}~\bibnamefont
  {Weis}}, \bibinfo {author} {\bibfnamefont {R.}~\bibnamefont {Rivi{\`e}re}},
  \bibinfo {author} {\bibfnamefont {S.}~\bibnamefont {Del{\'e}glise}}, \bibinfo
  {author} {\bibfnamefont {E.}~\bibnamefont {Gavartin}}, \bibinfo {author}
  {\bibfnamefont {O.}~\bibnamefont {Arcizet}}, \bibinfo {author} {\bibfnamefont
  {A.}~\bibnamefont {Schliesser}}, \ and\ \bibinfo {author} {\bibfnamefont
  {T.~J.}\ \bibnamefont {Kippenberg}},\ }\bibfield  {title} {\enquote {\bibinfo
  {title} {Optomechanically induced transparency},}\ }\href@noop {} {\bibfield
  {journal} {\bibinfo  {journal} {Science}\ }\textbf {\bibinfo {volume}
  {330}},\ \bibinfo {pages} {1520--1523} (\bibinfo {year} {2010})}\BibitemShut
  {NoStop}%
\bibitem [{\citenamefont {Kim}\ \emph {et~al.}(2015)\citenamefont {Kim},
  \citenamefont {Kuzyk}, \citenamefont {Han}, \citenamefont {Wang},\ and\
  \citenamefont {Bahl}}]{kim2015non}%
  \BibitemOpen
  \bibfield  {author} {\bibinfo {author} {\bibfnamefont {J.}~\bibnamefont
  {Kim}}, \bibinfo {author} {\bibfnamefont {M.~C.}\ \bibnamefont {Kuzyk}},
  \bibinfo {author} {\bibfnamefont {K.}~\bibnamefont {Han}}, \bibinfo {author}
  {\bibfnamefont {H.}~\bibnamefont {Wang}}, \ and\ \bibinfo {author}
  {\bibfnamefont {G.}~\bibnamefont {Bahl}},\ }\bibfield  {title} {\enquote
  {\bibinfo {title} {Non-reciprocal {B}rillouin scattering induced
  transparency},}\ }\href@noop {} {\bibfield  {journal} {\bibinfo  {journal}
  {Nature Physics}\ }\textbf {\bibinfo {volume} {11}},\ \bibinfo {pages}
  {275--280} (\bibinfo {year} {2015})}\BibitemShut {NoStop}%
\bibitem [{\citenamefont {Bahl}\ \emph {et~al.}(2012)\citenamefont {Bahl},
  \citenamefont {Tomes}, \citenamefont {Marquardt},\ and\ \citenamefont
  {Carmon}}]{bahl2012observation}%
  \BibitemOpen
  \bibfield  {author} {\bibinfo {author} {\bibfnamefont {G.}~\bibnamefont
  {Bahl}}, \bibinfo {author} {\bibfnamefont {M.}~\bibnamefont {Tomes}},
  \bibinfo {author} {\bibfnamefont {F.}~\bibnamefont {Marquardt}}, \ and\
  \bibinfo {author} {\bibfnamefont {T.}~\bibnamefont {Carmon}},\ }\bibfield
  {title} {\enquote {\bibinfo {title} {Observation of spontaneous {B}rillouin
  cooling},}\ }\href@noop {} {\bibfield  {journal} {\bibinfo  {journal} {Nature
  Physics}\ }\textbf {\bibinfo {volume} {8}},\ \bibinfo {pages} {203--207}
  (\bibinfo {year} {2012})}\BibitemShut {NoStop}%
\bibitem [{\citenamefont {Chan}\ \emph {et~al.}(2011)\citenamefont {Chan},
  \citenamefont {Alegre}, \citenamefont {Safavi-Naeini}, \citenamefont {Hill},
  \citenamefont {Krause}, \citenamefont {Gr{\"o}blacher}, \citenamefont
  {Aspelmeyer},\ and\ \citenamefont {Painter}}]{chan2011laser}%
  \BibitemOpen
  \bibfield  {author} {\bibinfo {author} {\bibfnamefont {J.}~\bibnamefont
  {Chan}}, \bibinfo {author} {\bibfnamefont {T.~M.}\ \bibnamefont {Alegre}},
  \bibinfo {author} {\bibfnamefont {A.~H.}\ \bibnamefont {Safavi-Naeini}},
  \bibinfo {author} {\bibfnamefont {J.~T.}\ \bibnamefont {Hill}}, \bibinfo
  {author} {\bibfnamefont {A.}~\bibnamefont {Krause}}, \bibinfo {author}
  {\bibfnamefont {S.}~\bibnamefont {Gr{\"o}blacher}}, \bibinfo {author}
  {\bibfnamefont {M.}~\bibnamefont {Aspelmeyer}}, \ and\ \bibinfo {author}
  {\bibfnamefont {O.}~\bibnamefont {Painter}},\ }\bibfield  {title} {\enquote
  {\bibinfo {title} {Laser cooling of a nanomechanical oscillator into its
  quantum ground state},}\ }\href@noop {} {\bibfield  {journal} {\bibinfo
  {journal} {Nature}\ }\textbf {\bibinfo {volume} {478}},\ \bibinfo {pages}
  {89--92} (\bibinfo {year} {2011})}\BibitemShut {NoStop}%
\bibitem [{\citenamefont {Teufel}\ \emph {et~al.}(2011)\citenamefont {Teufel},
  \citenamefont {Donner}, \citenamefont {Li}, \citenamefont {Harlow},
  \citenamefont {Allman}, \citenamefont {Cicak}, \citenamefont {Sirois},
  \citenamefont {Whittaker}, \citenamefont {Lehnert},\ and\ \citenamefont
  {Simmonds}}]{teufel2011sideband}%
  \BibitemOpen
  \bibfield  {author} {\bibinfo {author} {\bibfnamefont {J.}~\bibnamefont
  {Teufel}}, \bibinfo {author} {\bibfnamefont {T.}~\bibnamefont {Donner}},
  \bibinfo {author} {\bibfnamefont {D.}~\bibnamefont {Li}}, \bibinfo {author}
  {\bibfnamefont {J.}~\bibnamefont {Harlow}}, \bibinfo {author} {\bibfnamefont
  {M.}~\bibnamefont {Allman}}, \bibinfo {author} {\bibfnamefont
  {K.}~\bibnamefont {Cicak}}, \bibinfo {author} {\bibfnamefont
  {A.}~\bibnamefont {Sirois}}, \bibinfo {author} {\bibfnamefont {J.~D.}\
  \bibnamefont {Whittaker}}, \bibinfo {author} {\bibfnamefont {K.}~\bibnamefont
  {Lehnert}}, \ and\ \bibinfo {author} {\bibfnamefont {R.~W.}\ \bibnamefont
  {Simmonds}},\ }\bibfield  {title} {\enquote {\bibinfo {title} {Sideband
  cooling of micromechanical motion to the quantum ground state},}\ }\href@noop
  {} {\bibfield  {journal} {\bibinfo  {journal} {Nature}\ }\textbf {\bibinfo
  {volume} {475}},\ \bibinfo {pages} {359--363} (\bibinfo {year}
  {2011})}\BibitemShut {NoStop}%
\bibitem [{\citenamefont {Peano}\ \emph {et~al.}(2015)\citenamefont {Peano},
  \citenamefont {Brendel}, \citenamefont {Schmidt},\ and\ \citenamefont
  {Marquardt}}]{peano2015topological}%
  \BibitemOpen
  \bibfield  {author} {\bibinfo {author} {\bibfnamefont {V.}~\bibnamefont
  {Peano}}, \bibinfo {author} {\bibfnamefont {C.}~\bibnamefont {Brendel}},
  \bibinfo {author} {\bibfnamefont {M.}~\bibnamefont {Schmidt}}, \ and\
  \bibinfo {author} {\bibfnamefont {F.}~\bibnamefont {Marquardt}},\ }\bibfield
  {title} {\enquote {\bibinfo {title} {Topological phases of sound and
  light},}\ }\href@noop {} {\bibfield  {journal} {\bibinfo  {journal} {Physical
  Review X}\ }\textbf {\bibinfo {volume} {5}},\ \bibinfo {pages} {031011}
  (\bibinfo {year} {2015})}\BibitemShut {NoStop}%
\bibitem [{\citenamefont {Schmidt}\ \emph {et~al.}(2015)\citenamefont
  {Schmidt}, \citenamefont {Ke{\ss}ler}, \citenamefont {Peano}, \citenamefont
  {Painter},\ and\ \citenamefont {Marquardt}}]{schmidt2015optomechanical}%
  \BibitemOpen
  \bibfield  {author} {\bibinfo {author} {\bibfnamefont {M.}~\bibnamefont
  {Schmidt}}, \bibinfo {author} {\bibfnamefont {S.}~\bibnamefont {Ke{\ss}ler}},
  \bibinfo {author} {\bibfnamefont {V.}~\bibnamefont {Peano}}, \bibinfo
  {author} {\bibfnamefont {O.}~\bibnamefont {Painter}}, \ and\ \bibinfo
  {author} {\bibfnamefont {F.}~\bibnamefont {Marquardt}},\ }\bibfield  {title}
  {\enquote {\bibinfo {title} {Optomechanical creation of magnetic fields for
  photons on a lattice},}\ }\href@noop {} {\bibfield  {journal} {\bibinfo
  {journal} {Optica}\ }\textbf {\bibinfo {volume} {2}},\ \bibinfo {pages}
  {635--641} (\bibinfo {year} {2015})}\BibitemShut {NoStop}%
\bibitem [{\citenamefont {Kittlaus}, \citenamefont {Shin},\ and\ \citenamefont
  {Rakich}(2016)}]{kittlaus2015large}%
  \BibitemOpen
  \bibfield  {author} {\bibinfo {author} {\bibfnamefont {E.~A.}\ \bibnamefont
  {Kittlaus}}, \bibinfo {author} {\bibfnamefont {H.}~\bibnamefont {Shin}}, \
  and\ \bibinfo {author} {\bibfnamefont {P.~T.}\ \bibnamefont {Rakich}},\
  }\bibfield  {title} {\enquote {\bibinfo {title} {Large {B}rillouin
  amplification in silicon},}\ }\href@noop {} {\bibfield  {journal} {\bibinfo
  {journal} {Nature Photonics}\ } (\bibinfo {year} {2016})}\BibitemShut
  {NoStop}%
\bibitem [{\citenamefont {Arcizet}\ \emph {et~al.}(2006)\citenamefont
  {Arcizet}, \citenamefont {Cohadon}, \citenamefont {Briant}, \citenamefont
  {Pinard},\ and\ \citenamefont {Heidmann}}]{arcizet2006radiation}%
  \BibitemOpen
  \bibfield  {author} {\bibinfo {author} {\bibfnamefont {O.}~\bibnamefont
  {Arcizet}}, \bibinfo {author} {\bibfnamefont {P.-F.}\ \bibnamefont
  {Cohadon}}, \bibinfo {author} {\bibfnamefont {T.}~\bibnamefont {Briant}},
  \bibinfo {author} {\bibfnamefont {M.}~\bibnamefont {Pinard}}, \ and\ \bibinfo
  {author} {\bibfnamefont {A.}~\bibnamefont {Heidmann}},\ }\bibfield  {title}
  {\enquote {\bibinfo {title} {Radiation-pressure cooling and optomechanical
  instability of a micromirror},}\ }\href@noop {} {\bibfield  {journal}
  {\bibinfo  {journal} {Nature}\ }\textbf {\bibinfo {volume} {444}},\ \bibinfo
  {pages} {71--74} (\bibinfo {year} {2006})}\BibitemShut {NoStop}%
\bibitem [{\citenamefont {Kippenberg}\ and\ \citenamefont
  {Vahala}(2008)}]{kippenberg2008cavity}%
  \BibitemOpen
  \bibfield  {author} {\bibinfo {author} {\bibfnamefont {T.~J.}\ \bibnamefont
  {Kippenberg}}\ and\ \bibinfo {author} {\bibfnamefont {K.~J.}\ \bibnamefont
  {Vahala}},\ }\bibfield  {title} {\enquote {\bibinfo {title} {Cavity
  optomechanics: back-action at the mesoscale},}\ }\href@noop {} {\bibfield
  {journal} {\bibinfo  {journal} {science}\ }\textbf {\bibinfo {volume}
  {321}},\ \bibinfo {pages} {1172--1176} (\bibinfo {year} {2008})}\BibitemShut
  {NoStop}%
\bibitem [{\citenamefont {Byrnes}\ \emph {et~al.}(2012)\citenamefont {Byrnes},
  \citenamefont {Pant}, \citenamefont {Li}, \citenamefont {Choi}, \citenamefont
  {Poulton}, \citenamefont {Fan}, \citenamefont {Madden}, \citenamefont
  {Luther-Davies},\ and\ \citenamefont {Eggleton}}]{byrnes2012photonic}%
  \BibitemOpen
  \bibfield  {author} {\bibinfo {author} {\bibfnamefont {A.}~\bibnamefont
  {Byrnes}}, \bibinfo {author} {\bibfnamefont {R.}~\bibnamefont {Pant}},
  \bibinfo {author} {\bibfnamefont {E.}~\bibnamefont {Li}}, \bibinfo {author}
  {\bibfnamefont {D.-Y.}\ \bibnamefont {Choi}}, \bibinfo {author}
  {\bibfnamefont {C.~G.}\ \bibnamefont {Poulton}}, \bibinfo {author}
  {\bibfnamefont {S.}~\bibnamefont {Fan}}, \bibinfo {author} {\bibfnamefont
  {S.}~\bibnamefont {Madden}}, \bibinfo {author} {\bibfnamefont
  {B.}~\bibnamefont {Luther-Davies}}, \ and\ \bibinfo {author} {\bibfnamefont
  {B.~J.}\ \bibnamefont {Eggleton}},\ }\bibfield  {title} {\enquote {\bibinfo
  {title} {Photonic chip based tunable and reconfigurable narrowband microwave
  photonic filter using stimulated {B}rillouin scattering},}\ }\href@noop {}
  {\bibfield  {journal} {\bibinfo  {journal} {Optics Express}\ }\textbf
  {\bibinfo {volume} {20}},\ \bibinfo {pages} {18836--18845} (\bibinfo {year}
  {2012})}\BibitemShut {NoStop}%
\bibitem [{\citenamefont {Verhagen}\ \emph {et~al.}(2012)\citenamefont
  {Verhagen}, \citenamefont {Del{\'e}glise}, \citenamefont {Weis},
  \citenamefont {Schliesser},\ and\ \citenamefont
  {Kippenberg}}]{verhagen2012quantum}%
  \BibitemOpen
  \bibfield  {author} {\bibinfo {author} {\bibfnamefont {E.}~\bibnamefont
  {Verhagen}}, \bibinfo {author} {\bibfnamefont {S.}~\bibnamefont
  {Del{\'e}glise}}, \bibinfo {author} {\bibfnamefont {S.}~\bibnamefont {Weis}},
  \bibinfo {author} {\bibfnamefont {A.}~\bibnamefont {Schliesser}}, \ and\
  \bibinfo {author} {\bibfnamefont {T.~J.}\ \bibnamefont {Kippenberg}},\
  }\bibfield  {title} {\enquote {\bibinfo {title} {Quantum-coherent coupling of
  a mechanical oscillator to an optical cavity mode},}\ }\href@noop {}
  {\bibfield  {journal} {\bibinfo  {journal} {Nature}\ }\textbf {\bibinfo
  {volume} {482}},\ \bibinfo {pages} {63--67} (\bibinfo {year}
  {2012})}\BibitemShut {NoStop}%
\bibitem [{\citenamefont {Palomaki}\ \emph {et~al.}(2013)\citenamefont
  {Palomaki}, \citenamefont {Harlow}, \citenamefont {Teufel}, \citenamefont
  {Simmonds},\ and\ \citenamefont {Lehnert}}]{palomaki2013coherent}%
  \BibitemOpen
  \bibfield  {author} {\bibinfo {author} {\bibfnamefont {T.}~\bibnamefont
  {Palomaki}}, \bibinfo {author} {\bibfnamefont {J.}~\bibnamefont {Harlow}},
  \bibinfo {author} {\bibfnamefont {J.}~\bibnamefont {Teufel}}, \bibinfo
  {author} {\bibfnamefont {R.}~\bibnamefont {Simmonds}}, \ and\ \bibinfo
  {author} {\bibfnamefont {K.}~\bibnamefont {Lehnert}},\ }\bibfield  {title}
  {\enquote {\bibinfo {title} {Coherent state transfer between itinerant
  microwave fields and a mechanical oscillator},}\ }\href@noop {} {\bibfield
  {journal} {\bibinfo  {journal} {Nature}\ }\textbf {\bibinfo {volume} {495}},\
  \bibinfo {pages} {210--214} (\bibinfo {year} {2013})}\BibitemShut {NoStop}%
\bibitem [{\citenamefont {Bochmann}\ \emph {et~al.}(2013)\citenamefont
  {Bochmann}, \citenamefont {Vainsencher}, \citenamefont {Awschalom},\ and\
  \citenamefont {Cleland}}]{bochmann2013nanomechanical}%
  \BibitemOpen
  \bibfield  {author} {\bibinfo {author} {\bibfnamefont {J.}~\bibnamefont
  {Bochmann}}, \bibinfo {author} {\bibfnamefont {A.}~\bibnamefont
  {Vainsencher}}, \bibinfo {author} {\bibfnamefont {D.~D.}\ \bibnamefont
  {Awschalom}}, \ and\ \bibinfo {author} {\bibfnamefont {A.~N.}\ \bibnamefont
  {Cleland}},\ }\bibfield  {title} {\enquote {\bibinfo {title} {Nanomechanical
  coupling between microwave and optical photons},}\ }\href@noop {} {\bibfield
  {journal} {\bibinfo  {journal} {Nature Physics}\ }\textbf {\bibinfo {volume}
  {9}},\ \bibinfo {pages} {712--716} (\bibinfo {year} {2013})}\BibitemShut
  {NoStop}%
\bibitem [{\citenamefont {Lu}\ \emph {et~al.}(2013)\citenamefont {Lu},
  \citenamefont {Galipeau}, \citenamefont {Mouthaan}, \citenamefont {Briot},\
  and\ \citenamefont {Abbott}}]{lu2013reconfigurable}%
  \BibitemOpen
  \bibfield  {author} {\bibinfo {author} {\bibfnamefont {X.}~\bibnamefont
  {Lu}}, \bibinfo {author} {\bibfnamefont {J.}~\bibnamefont {Galipeau}},
  \bibinfo {author} {\bibfnamefont {K.}~\bibnamefont {Mouthaan}}, \bibinfo
  {author} {\bibfnamefont {E.~H.}\ \bibnamefont {Briot}}, \ and\ \bibinfo
  {author} {\bibfnamefont {B.}~\bibnamefont {Abbott}},\ }\bibfield  {title}
  {\enquote {\bibinfo {title} {Reconfigurable multiband saw filters for lte
  applications},}\ }in\ \href@noop {} {\emph {\bibinfo {booktitle} {Radio and
  Wireless Symposium (RWS), 2013 IEEE}}}\ (\bibinfo {organization} {IEEE},\
  \bibinfo {year} {2013})\ pp.\ \bibinfo {pages} {253--255}\BibitemShut
  {NoStop}%
\bibitem [{\citenamefont {Psychogiou}\ \emph {et~al.}(2015)\citenamefont
  {Psychogiou}, \citenamefont {G{\'o}mez-Garc{\i}}, \citenamefont {Peroulis}
  \emph {et~al.}}]{psychogiou2015acoustic}%
  \BibitemOpen
  \bibfield  {author} {\bibinfo {author} {\bibfnamefont {D.}~\bibnamefont
  {Psychogiou}}, \bibinfo {author} {\bibfnamefont {R.}~\bibnamefont
  {G{\'o}mez-Garc{\i}}}, \bibinfo {author} {\bibfnamefont {D.}~\bibnamefont
  {Peroulis}},  \emph {et~al.},\ }\bibfield  {title} {\enquote {\bibinfo
  {title} {Acoustic wave resonator-based absorptive bandstop filters with
  ultra-narrow bandwidth},}\ }\href@noop {} {\bibfield  {journal} {\bibinfo
  {journal} {IEEE Microwave and Wireless Components Letters}\ }\textbf
  {\bibinfo {volume} {25}},\ \bibinfo {pages} {570--572} (\bibinfo {year}
  {2015})}\BibitemShut {NoStop}%
\bibitem [{\citenamefont {Clark}\ \emph {et~al.}(2005)\citenamefont {Clark},
  \citenamefont {Hsu}, \citenamefont {Abdelmoneum},\ and\ \citenamefont
  {Nguyen}}]{clark2005high}%
  \BibitemOpen
  \bibfield  {author} {\bibinfo {author} {\bibfnamefont {J.~R.}\ \bibnamefont
  {Clark}}, \bibinfo {author} {\bibfnamefont {W.-T.}\ \bibnamefont {Hsu}},
  \bibinfo {author} {\bibfnamefont {M.~A.}\ \bibnamefont {Abdelmoneum}}, \ and\
  \bibinfo {author} {\bibfnamefont {C.-C.}\ \bibnamefont {Nguyen}},\ }\bibfield
   {title} {\enquote {\bibinfo {title} {High-{Q} {UHF} micromechanical
  radial-contour mode disk resonators},}\ }\href@noop {} {\bibfield  {journal}
  {\bibinfo  {journal} {Journal of Microelectromechanical Systems}\ }\textbf
  {\bibinfo {volume} {14}},\ \bibinfo {pages} {1298--1310} (\bibinfo {year}
  {2005})}\BibitemShut {NoStop}%
\bibitem [{\citenamefont {Weinstein}\ and\ \citenamefont
  {Bhave}(2010)}]{weinstein2010resonant}%
  \BibitemOpen
  \bibfield  {author} {\bibinfo {author} {\bibfnamefont {D.}~\bibnamefont
  {Weinstein}}\ and\ \bibinfo {author} {\bibfnamefont {S.~A.}\ \bibnamefont
  {Bhave}},\ }\bibfield  {title} {\enquote {\bibinfo {title} {The resonant body
  transistor},}\ }\href@noop {} {\bibfield  {journal} {\bibinfo  {journal}
  {Nano Letters}\ }\textbf {\bibinfo {volume} {10}},\ \bibinfo {pages}
  {1234--1237} (\bibinfo {year} {2010})}\BibitemShut {NoStop}%
\bibitem [{\citenamefont {Petersson}\ and\ \citenamefont
  {Sj{\"o}green}(2015)}]{petersson2015wave}%
  \BibitemOpen
  \bibfield  {author} {\bibinfo {author} {\bibfnamefont {N.~A.}\ \bibnamefont
  {Petersson}}\ and\ \bibinfo {author} {\bibfnamefont {B.}~\bibnamefont
  {Sj{\"o}green}},\ }\bibfield  {title} {\enquote {\bibinfo {title} {Wave
  propagation in anisotropic elastic materials and curvilinear coordinates
  using a summation-by-parts finite difference method},}\ }\href@noop {}
  {\bibfield  {journal} {\bibinfo  {journal} {Journal of Computational
  Physics}\ }\textbf {\bibinfo {volume} {299}},\ \bibinfo {pages} {820--841}
  (\bibinfo {year} {2015})}\BibitemShut {NoStop}%
\bibitem [{\citenamefont {Gao}\ \emph {et~al.}(2015)\citenamefont {Gao},
  \citenamefont {Fu}, \citenamefont {Gibson}, \citenamefont {Chung},\ and\
  \citenamefont {Efendiev}}]{gao2015generalized}%
  \BibitemOpen
  \bibfield  {author} {\bibinfo {author} {\bibfnamefont {K.}~\bibnamefont
  {Gao}}, \bibinfo {author} {\bibfnamefont {S.}~\bibnamefont {Fu}}, \bibinfo
  {author} {\bibfnamefont {R.~L.}\ \bibnamefont {Gibson}}, \bibinfo {author}
  {\bibfnamefont {E.~T.}\ \bibnamefont {Chung}}, \ and\ \bibinfo {author}
  {\bibfnamefont {Y.}~\bibnamefont {Efendiev}},\ }\bibfield  {title} {\enquote
  {\bibinfo {title} {Generalized multiscale finite-element method (gmsfem) for
  elastic wave propagation in heterogeneous, anisotropic media},}\ }\href@noop
  {} {\bibfield  {journal} {\bibinfo  {journal} {Journal of Computational
  Physics}\ }\textbf {\bibinfo {volume} {295}},\ \bibinfo {pages} {161--188}
  (\bibinfo {year} {2015})}\BibitemShut {NoStop}%
\bibitem [{\citenamefont {{COMSOL
  Multiphysics}}(2012)}]{multiphysics2012comsol}%
  \BibitemOpen
  \bibfield  {author} {\bibinfo {author} {\bibnamefont {{COMSOL
  Multiphysics}}},\ }\href@noop {} {\emph {\bibinfo {title} {version 4.3a}}}\
  (\bibinfo  {publisher} {COMSOL Inc.},\ \bibinfo {address} {Burlington,
  Massachusetts},\ \bibinfo {year} {2012})\BibitemShut {NoStop}%
\bibitem [{\citenamefont {Gavri{\'c}}(1995)}]{gavric1995computation}%
  \BibitemOpen
  \bibfield  {author} {\bibinfo {author} {\bibfnamefont {L.}~\bibnamefont
  {Gavri{\'c}}},\ }\bibfield  {title} {\enquote {\bibinfo {title} {Computation
  of propagative waves in free rail using a finite element technique},}\
  }\href@noop {} {\bibfield  {journal} {\bibinfo  {journal} {Journal of Sound
  and Vibration}\ }\textbf {\bibinfo {volume} {185}},\ \bibinfo {pages}
  {531--543} (\bibinfo {year} {1995})}\BibitemShut {NoStop}%
\bibitem [{\citenamefont {Wilcox}\ \emph {et~al.}(2002)\citenamefont {Wilcox},
  \citenamefont {Evans}, \citenamefont {Diligent}, \citenamefont {Lowe},\ and\
  \citenamefont {Cawley}}]{wilcox2002dispersion}%
  \BibitemOpen
  \bibfield  {author} {\bibinfo {author} {\bibfnamefont {P.}~\bibnamefont
  {Wilcox}}, \bibinfo {author} {\bibfnamefont {M.}~\bibnamefont {Evans}},
  \bibinfo {author} {\bibfnamefont {O.}~\bibnamefont {Diligent}}, \bibinfo
  {author} {\bibfnamefont {M.}~\bibnamefont {Lowe}}, \ and\ \bibinfo {author}
  {\bibfnamefont {P.}~\bibnamefont {Cawley}},\ }\bibfield  {title} {\enquote
  {\bibinfo {title} {Dispersion and excitability of guided acoustic waves in
  isotropic beams with arbitrary cross section},}\ }in\ \href@noop {} {\emph
  {\bibinfo {booktitle} {AIP Conference Proceedings}}},\ Vol.\ \bibinfo
  {volume} {615}\ (\bibinfo {organization} {AIP},\ \bibinfo {year} {2002})\
  pp.\ \bibinfo {pages} {203--210}\BibitemShut {NoStop}%
\bibitem [{\citenamefont {Castaings}\ and\ \citenamefont
  {Lowe}(2008)}]{castaings2008finite}%
  \BibitemOpen
  \bibfield  {author} {\bibinfo {author} {\bibfnamefont {M.}~\bibnamefont
  {Castaings}}\ and\ \bibinfo {author} {\bibfnamefont {M.}~\bibnamefont
  {Lowe}},\ }\bibfield  {title} {\enquote {\bibinfo {title} {Finite element
  model for waves guided along solid systems of arbitrary section coupled to
  infinite solid media},}\ }\href@noop {} {\bibfield  {journal} {\bibinfo
  {journal} {The Journal of the Acoustical Society of America}\ }\textbf
  {\bibinfo {volume} {123}},\ \bibinfo {pages} {696--708} (\bibinfo {year}
  {2008})}\BibitemShut {NoStop}%
\bibitem [{\citenamefont {Bartoli}\ \emph {et~al.}(2006)\citenamefont
  {Bartoli}, \citenamefont {Marzani}, \citenamefont {di~Scalea},\ and\
  \citenamefont {Viola}}]{bartoli2006modeling}%
  \BibitemOpen
  \bibfield  {author} {\bibinfo {author} {\bibfnamefont {I.}~\bibnamefont
  {Bartoli}}, \bibinfo {author} {\bibfnamefont {A.}~\bibnamefont {Marzani}},
  \bibinfo {author} {\bibfnamefont {F.~L.}\ \bibnamefont {di~Scalea}}, \ and\
  \bibinfo {author} {\bibfnamefont {E.}~\bibnamefont {Viola}},\ }\bibfield
  {title} {\enquote {\bibinfo {title} {Modeling wave propagation in damped
  waveguides of arbitrary cross-section},}\ }\href@noop {} {\bibfield
  {journal} {\bibinfo  {journal} {Journal of Sound and Vibration}\ }\textbf
  {\bibinfo {volume} {295}},\ \bibinfo {pages} {685--707} (\bibinfo {year}
  {2006})}\BibitemShut {NoStop}%
\bibitem [{\citenamefont {Yee}\ \emph {et~al.}(1966)\citenamefont {Yee} \emph
  {et~al.}}]{yee1966numerical}%
  \BibitemOpen
  \bibfield  {author} {\bibinfo {author} {\bibfnamefont {K.~S.}\ \bibnamefont
  {Yee}} \emph {et~al.},\ }\bibfield  {title} {\enquote {\bibinfo {title}
  {Numerical solution of initial boundary value problems involving maxwell’s
  equations in isotropic media},}\ }\href@noop {} {\bibfield  {journal}
  {\bibinfo  {journal} {IEEE Trans. Antennas Propag.}\ }\textbf {\bibinfo
  {volume} {14}},\ \bibinfo {pages} {302--307} (\bibinfo {year}
  {1966})}\BibitemShut {NoStop}%
\bibitem [{\citenamefont {Madariaga}(1976)}]{madariaga1976dynamics}%
  \BibitemOpen
  \bibfield  {author} {\bibinfo {author} {\bibfnamefont {R.}~\bibnamefont
  {Madariaga}},\ }\bibfield  {title} {\enquote {\bibinfo {title} {Dynamics of
  an expanding circular fault},}\ }\href@noop {} {\bibfield  {journal}
  {\bibinfo  {journal} {Bulletin of the Seismological Society of America}\
  }\textbf {\bibinfo {volume} {66}},\ \bibinfo {pages} {639--666} (\bibinfo
  {year} {1976})}\BibitemShut {NoStop}%
\bibitem [{\citenamefont {Etemadsaeed}\ \emph {et~al.}(2016)\citenamefont
  {Etemadsaeed}, \citenamefont {Moczo}, \citenamefont {Kristek}, \citenamefont
  {Ansari},\ and\ \citenamefont {Kristekova}}]{etemadsaeed2016no}%
  \BibitemOpen
  \bibfield  {author} {\bibinfo {author} {\bibfnamefont {L.}~\bibnamefont
  {Etemadsaeed}}, \bibinfo {author} {\bibfnamefont {P.}~\bibnamefont {Moczo}},
  \bibinfo {author} {\bibfnamefont {J.}~\bibnamefont {Kristek}}, \bibinfo
  {author} {\bibfnamefont {A.}~\bibnamefont {Ansari}}, \ and\ \bibinfo {author}
  {\bibfnamefont {M.}~\bibnamefont {Kristekova}},\ }\bibfield  {title}
  {\enquote {\bibinfo {title} {A no-cost improved velocity--stress
  staggered-grid finite-difference scheme for modelling seismic wave
  propagation},}\ }\href@noop {} {\bibfield  {journal} {\bibinfo  {journal}
  {Geophysical Journal International}\ }\textbf {\bibinfo {volume} {207}},\
  \bibinfo {pages} {481--511} (\bibinfo {year} {2016})}\BibitemShut {NoStop}%
\bibitem [{\citenamefont {{MATLAB}}(2015)}]{matlab}%
  \BibitemOpen
  \bibfield  {author} {\bibinfo {author} {\bibnamefont {{MATLAB}}},\
  }\href@noop {} {\emph {\bibinfo {title} {version R2015a}}}\ (\bibinfo
  {publisher} {The MathWorks Inc.},\ \bibinfo {address} {Natick,
  Massachusetts},\ \bibinfo {year} {2015})\BibitemShut {NoStop}%
\bibitem [{\citenamefont {Dostart}\ and\ \citenamefont
  {Popovi\'c}(2016)}]{mathworks}%
  \BibitemOpen
  \bibfield  {author} {\bibinfo {author} {\bibfnamefont {N.}~\bibnamefont
  {Dostart}}\ and\ \bibinfo {author} {\bibfnamefont {M.}~\bibnamefont
  {Popovi\'c}},\ }\href
  {https://www.mathworks.com/matlabcentral/fileexchange/58729-elastic-mode-solver}
  {\enquote {\bibinfo {title} {Matlab central file exchange: Elastic mode
  solver},}\ } (\bibinfo {year} {Mathworks, 2016}),\ \bibinfo {note} {[Online;
  accessed 14-April-2017]}\BibitemShut {NoStop}%
\bibitem [{\citenamefont {Snyder}\ and\ \citenamefont
  {Love}(2012)}]{snyder2012optical}%
  \BibitemOpen
  \bibfield  {author} {\bibinfo {author} {\bibfnamefont {A.~W.}\ \bibnamefont
  {Snyder}}\ and\ \bibinfo {author} {\bibfnamefont {J.~D.}\ \bibnamefont
  {Love}},\ }\href@noop {} {\emph {\bibinfo {title} {Optical Waveguide
  Theory}}}\ (\bibinfo  {publisher} {Springer Science \& Business Media},\
  \bibinfo {year} {2012})\BibitemShut {NoStop}%
\bibitem [{\citenamefont {Auld}(1990)}]{Auld1990}%
  \BibitemOpen
  \bibfield  {author} {\bibinfo {author} {\bibfnamefont {B.~A.}\ \bibnamefont
  {Auld}},\ }\href@noop {} {\emph {\bibinfo {title} {Acoustic Fields and Waves
  in Solids}}},\ \bibinfo {edition} {2nd}\ ed.\ (\bibinfo  {publisher}
  {Krieger},\ \bibinfo {address} {Malabar},\ \bibinfo {year}
  {1990})\BibitemShut {NoStop}%
\bibitem [{\citenamefont {Chew}(1994)}]{chew1994em}%
  \BibitemOpen
  \bibfield  {author} {\bibinfo {author} {\bibfnamefont {W.~C.}\ \bibnamefont
  {Chew}},\ }\bibfield  {title} {\enquote {\bibinfo {title} {Electromagnetic
  theory on a lattice},}\ }\href@noop {} {\bibfield  {journal} {\bibinfo
  {journal} {Journal of Applied Physics}\ }\textbf {\bibinfo {volume} {75}},\
  \bibinfo {pages} {4843--4850} (\bibinfo {year} {1994})}\BibitemShut {NoStop}%
\bibitem [{\citenamefont {Joannopoulos}\ \emph {et~al.}(2011)\citenamefont
  {Joannopoulos}, \citenamefont {Johnson}, \citenamefont {Winn},\ and\
  \citenamefont {Meade}}]{joannopoulos2011photonic}%
  \BibitemOpen
  \bibfield  {author} {\bibinfo {author} {\bibfnamefont {J.~D.}\ \bibnamefont
  {Joannopoulos}}, \bibinfo {author} {\bibfnamefont {S.~G.}\ \bibnamefont
  {Johnson}}, \bibinfo {author} {\bibfnamefont {J.~N.}\ \bibnamefont {Winn}}, \
  and\ \bibinfo {author} {\bibfnamefont {R.~D.}\ \bibnamefont {Meade}},\
  }\href@noop {} {\emph {\bibinfo {title} {Photonic Crystals: Molding the Flow
  of Light}}}\ (\bibinfo  {publisher} {Princeton university press},\ \bibinfo
  {year} {2011})\BibitemShut {NoStop}%
\bibitem [{\citenamefont {Chew}\ and\ \citenamefont
  {Liu}(1996)}]{chew1996perfectly}%
  \BibitemOpen
  \bibfield  {author} {\bibinfo {author} {\bibfnamefont {W.~C.}\ \bibnamefont
  {Chew}}\ and\ \bibinfo {author} {\bibfnamefont {Q.~H.}\ \bibnamefont {Liu}},\
  }\bibfield  {title} {\enquote {\bibinfo {title} {Perfectly matched layers for
  elastodynamics: a new absorbing boundary condition},}\ }\href@noop {}
  {\bibfield  {journal} {\bibinfo  {journal} {Journal of Computational
  Acoustics}\ }\textbf {\bibinfo {volume} {4}},\ \bibinfo {pages} {341--359}
  (\bibinfo {year} {1996})}\BibitemShut {NoStop}%
\bibitem [{\citenamefont {Chew}\ and\ \citenamefont
  {Weedon}(1994)}]{chew19943d}%
  \BibitemOpen
  \bibfield  {author} {\bibinfo {author} {\bibfnamefont {W.~C.}\ \bibnamefont
  {Chew}}\ and\ \bibinfo {author} {\bibfnamefont {W.~H.}\ \bibnamefont
  {Weedon}},\ }\bibfield  {title} {\enquote {\bibinfo {title} {A 3{D} perfectly
  matched medium from modified {M}axwell's equations with stretched
  coordinates},}\ }\href@noop {} {\bibfield  {journal} {\bibinfo  {journal}
  {Microwave and Optical Technology Letters}\ }\textbf {\bibinfo {volume}
  {7}},\ \bibinfo {pages} {599--604} (\bibinfo {year} {1994})}\BibitemShut
  {NoStop}%
\bibitem [{\citenamefont {Berenger}(1994)}]{berenger1994perfectly}%
  \BibitemOpen
  \bibfield  {author} {\bibinfo {author} {\bibfnamefont {J.-P.}\ \bibnamefont
  {Berenger}},\ }\bibfield  {title} {\enquote {\bibinfo {title} {A perfectly
  matched layer for the absorption of electromagnetic waves},}\ }\href@noop {}
  {\bibfield  {journal} {\bibinfo  {journal} {Journal of computational
  physics}\ }\textbf {\bibinfo {volume} {114}},\ \bibinfo {pages} {185--200}
  (\bibinfo {year} {1994})}\BibitemShut {NoStop}%
\bibitem [{\citenamefont {Teixeira}\ and\ \citenamefont
  {Chew}(1998)}]{teixeira1998analytical}%
  \BibitemOpen
  \bibfield  {author} {\bibinfo {author} {\bibfnamefont {F.}~\bibnamefont
  {Teixeira}}\ and\ \bibinfo {author} {\bibfnamefont {W.}~\bibnamefont
  {Chew}},\ }\bibfield  {title} {\enquote {\bibinfo {title} {Analytical
  derivation of a conformal perfectly matched absorber for electromagnetic
  waves},}\ }\href@noop {} {\bibfield  {journal} {\bibinfo  {journal}
  {Microwave and Optical technology letters}\ }\textbf {\bibinfo {volume}
  {17}},\ \bibinfo {pages} {231--236} (\bibinfo {year} {1998})}\BibitemShut
  {NoStop}%
\bibitem [{\citenamefont {Popovi{\'c}}(2008)}]{popovic2008theory}%
  \BibitemOpen
  \bibfield  {author} {\bibinfo {author} {\bibfnamefont {M.}~\bibnamefont
  {Popovi{\'c}}},\ }\emph {\bibinfo {title} {Theory and design of
  high-index-contrast microphotonic circuits}},\ \href@noop {} {Ph.D. thesis},\
  \bibinfo  {school} {Massachusetts Institute of Technology} (\bibinfo {year}
  {2008})\BibitemShut {NoStop}%
\bibitem [{\citenamefont {Feng}\ \emph {et~al.}(2002)\citenamefont {Feng},
  \citenamefont {Zhou}, \citenamefont {Xu},\ and\ \citenamefont
  {Huang}}]{feng2002computation}%
  \BibitemOpen
  \bibfield  {author} {\bibinfo {author} {\bibfnamefont {N.-N.}\ \bibnamefont
  {Feng}}, \bibinfo {author} {\bibfnamefont {G.-R.}\ \bibnamefont {Zhou}},
  \bibinfo {author} {\bibfnamefont {C.}~\bibnamefont {Xu}}, \ and\ \bibinfo
  {author} {\bibfnamefont {W.-P.}\ \bibnamefont {Huang}},\ }\bibfield  {title}
  {\enquote {\bibinfo {title} {Computation of full-vector modes for bending
  waveguide using cylindrical perfectlymatched layers},}\ }\href@noop {}
  {\bibfield  {journal} {\bibinfo  {journal} {Journal of lightwave technology}\
  }\textbf {\bibinfo {volume} {20}},\ \bibinfo {pages} {1976} (\bibinfo {year}
  {2002})}\BibitemShut {NoStop}%
\end{thebibliography}
%

\end{document}